\documentclass[fleqn,usenatbib]{mnras}

\usepackage{newtxtext,newtxmath}
\usepackage[T1]{fontenc}

\DeclareRobustCommand{\VAN}[3]{#2}
\let\VANthebibliography\thebibliography
\def\thebibliography{\DeclareRobustCommand{\VAN}[3]{##3}\VANthebibliography}

\usepackage{graphicx}	% Including figure files
\usepackage{amsmath}	% Advanced maths commands

\title[Pristine Dwarf Galaxy Survey VII]{The Pristine Dwarf Galaxy Survey -- VII. The metallicity distributions of 12 Milky Way faint satellites}

\author[S. Taibi et al.]{
S.~Taibi,$^{1}$\thanks{E-mail: salvatore.taibi@epfl.ch}
P.~Jablonka,$^{1,2}$
N.~Longeard,$^{3}$
N.~F.~Martin,$^{4,5}$
Z.~Yuan,$^{6,7}$
G.~Battaglia,$^{8,9}$
A.~Ardern-Arentsen,$^{10}$
\newauthor
P.~C\^ot\'e,$^{11}$
J.~F.~Navarro,$^{12}$
R.~Sanchez-Janssen,$^{8,13}$
F.~Sestito,$^{14}$
E.~Starkenburg$^{15}$
and M.~Hirschmann$^{1}$
\\
$^{1}$Institute of Physics, Laboratory of Astrophysics, Ecole Polytechnique Fédérale de Lausanne (EPFL), 1290 Sauverny, Switzerland\\
$^{2}$LIRA, Observatoire de Paris, Université PSL, Sorbonne Université, Université Paris Cité, CY Cergy Paris Université, CNRS, F-92190 Meudon, France\\
$^{3}$IFCA, Instituto de Física de Cantabria (UC-CSIC), Av. de Los Castros s/n, E-39005 Santander, Spain\\
$^{4}$Observatoire astronomique de Strasbourg, Université de Strasbourg, CNRS, UMR 7550, F-67000 Strasbourg, France\\
$^{5}$Max-Planck-Institut für Astronomie, Königstuhl 17, D-69117 Heidelberg, Germany\\
$^{6}$School of Astronomy and Space Science, Nanjing University, Nanjing 210093, People’s Republic of China\\
$^{7}$Key Laboratory of Modern Astronomy and Astrophysics (Nanjing University), Ministry of Education, Nanjing 210093, People’s Republic of China\\
$^{8}$Instituto de Astrofísica de Canarias, Calle Vía Láctea s/n, E-38206 La Laguna, Santa Cruz de Tenerife, Spain\\
$^{9}$Universidad de La Laguna, Avda. Astrofísico Francisco Sánchez, E-38205 La Laguna, Santa Cruz de Tenerife, Spain\\
$^{10}$Institute of Astronomy, University of Cambridge, Madingley Rd, Cambridge CB3 0HA, UK\\
$^{11}$National Research Council of Canada, 5071 West Saanich Road, Victoria, BC, V9E 2E7, Canada\\
$^{12}$Dept. of Physics and Astronomy, University of Victoria, PO Box 3055, STN CSC, Victoria, BC V8W 3P6, Canada\\
$^{13}$Isaac Newton Group of Telescopes, E-38700 Santa Cruz de La Palma, Canary Islands, Spain\\
$^{14}$Centre for Astrophysics Research, Department of Physics, Astronomy and Mathematics, University of Hertfordshire, Hatfield AL10 9AB, UK\\
$^{15}$Kapteyn Astronomical Institute, University of Groningen, Landleven 12, 9747 AD Groningen, The Netherlands
}

\date{Accepted XXX. Received YYY; in original form ZZZ}

\pubyear{\the\year{}}

\begin{document}
\label{firstpage}
\pagerange{\pageref{firstpage}--\pageref{lastpage}}
\maketitle

\begin{abstract}

Spectroscopic studies of ultra-faint dwarf galaxies are typically limited to small samples of stars due to the scarcity of sufficiently bright targets. The small number statistics and possible presence of contaminants still hamper solid determinations of their metallicity distribution function.
In this work, we characterise the metallicity distributions of 12 Milky Way faint satellites, spanning the luminosity range $10^2 \lesssim L_V/L_\odot \lesssim 10^5$, by exploiting deep narrow-band \textit{CaHK} photometry from the \textit{Pristine} dwarf galaxy survey.
We combined narrow-band \textit{CaHK} photometry with deep broad-band $g$ and $r$ photometry from \citet{Munoz2018} and Pan-STARRS1, covering each system out to $5$--$8 \times R_h$ and reaching magnitudes as faint as $g_0\sim23$, in order to derive accurate stellar photometric metallicities. Membership probabilities were determined incorporating the available spatial, photometric, astrometric, and spectroscopic information, further refined using the derived photometric metallicities.
We identified 3\,917 probable member stars (with a membership probability $\mathcal{P} \geq 0.8$) across the 12 systems, more than doubling the numbers recovered by previous spectroscopic studies. We deliver complete metallicity distributions that yield robust average metallicities and dispersions previously inaccessible for most of the systems examined in this study.
We identify 170 candidate extremely metal-poor stars distributed across all systems, and confirm a departure from the linear luminosity-metallicity relation in the ultra-faint regime, with systems scattering around $\mathrm{[Fe/H]} \sim -2.3$~dex. 
Given the extensive mass coverage of our sample, we were able to investigate the presence of metallicity gradients, finding clear evidence of radial variations in massive systems, but none in the ultra-faint dwarfs within $2.5 \times R_h$.
The photometric strategy presented in this paper will continue to serve as an effective complement to future spectroscopic surveys.

\end{abstract}

\begin{keywords}
keyword1 -- keyword2 -- keyword3
\end{keywords}

%%%%%%%%%%%%%%%%%%%%%%%%%%%%%%%%%%%%%%%%%%%%%%%%%%

%%%%%%%%%%%%%%%%% BODY OF PAPER %%%%%%%%%%%%%%%%%%

\section{Introduction}
\label{sec:intro}

In the hierarchical formation of galaxies, the first systems emerge from the collapse of small dark matter halos, which then merge to form increasingly larger systems \citep[e.g.][]{Searle+Zinn1978,White+Rees1978,Bullock+Johnston2005, Helmi2020}. 
Our current understanding suggests that some of these building blocks did not undergo subsequent mergers and can still be observed today as dwarf galaxies. These fossils provide unique evidence of the conditions in the early Universe and are essential for testing the predictions of cosmological models at the smallest mass scales \citep[e.g.][]{Sales2022}.

The satellite dwarf galaxies of the Milky Way (MW) are particularly valuable laboratories. Their proximity allows them to be resolved into individual stars, enabling detailed studies of their dynamical and chemical properties. In particular, their metallicity distributions provide a record of the effects of star formation, gas accretion, and chemical enrichment throughout their evolutionary histories \citep[e.g.][]{Lin+Faber1983, Dekel+Silk1986, Kirby2013}. 
While the classical dwarf spheroidal galaxies (dSphs), with luminosities of $10^5$--$10^7\,L_\odot$, have been extensively investigated \citep[e.g.][]{Tolstoy2009, Kirby2013, Walker2023}, the characterisation of the smallest systems, known as ultra-faint dwarfs (UFDs, $L_V\lesssim 10^5\,L_\odot$), remains a major challenge.

Discovered in large numbers since the advent of wide-field photometric surveys \citep[e.g.][]{Willman2005a, Belokurov2006, Belokurov2007,Martin2007,Koposov2015}, UFDs are the oldest, most metal-poor and most dark-matter-dominated galaxies known \citep[e.g.][]{Simon+Geha2007,Laevens2015,Simon2019,Geha2026}.  Their star formation activity ended at or shortly after the reionisation epoch \citep[e.g.][]{Okamoto2012, Brown2014, Weisz2014}, limiting their chemical enrichment. Consequently, UFDs serve as ideal probes of the nucleosynthesis yields of the first generations of stars \citep[e.g.][]{Frebel+Norris2015, Simon2019}. Despite their relevance, the properties of UFDs remain poorly constrained, due to their faintness and the strong foreground contamination from the MW \citep[e.g.][]{Martin2007, Simon+Geha2007, Kirby2013}. 

Theoretical studies reconstructing the formation and evolution of UFDs using existing datasets face two key limitations \citep[see also the discussion in][]{Sanati2023}. On the one hand, the spectroscopic samples are heterogeneous and offer poor metallicity coverage. Consequently, the average metallicities calculated and subsequently used do not reflect the peaks of the metallicity distribution functions. On the other hand, the number of extremely metal-poor stars found in UFDs is surprisingly low and likely due to observational limitations. Conversely, some UFDs contain candidate stars with metallicities comparable to those of much more massive systems, leading to the suspicion of significant foreground contamination.

A particular issue in this context is the reproduction of observed scaling relations, such as the one between the luminosity and the average metallicity of a dwarf galaxy \citep[e.g.][]{Kirby2013}. Cosmological simulations produce UFDs that are systematically more metal-poor at a given luminosity than the extrapolated luminosity-metallicity relation obtained from brighter dwarf galaxies \citep[e.g.][]{Applebaum2021,Jeon2021,Sanati2023}. 
So far, the observed UFDs show considerable dispersion in the luminosity–metallicity plane, seemingly reaching a metallicity floor at $\sim-2.5$~dex \citep[e.g.][]{Fu2023,Geha2026b}. It remains unclear whether this floor is an intrinsic physical feature of UFDs or an artefact of sparse samples and foreground contamination. Based on the existing information, the proposed explanations range from tidal stripping \citep{Simon2019} to pre-enrichment of the interstellar medium \citep{Ahvazi2024, Rey2025arXiv}. It is undeniable that progress in our understanding of UFDs must rely on larger stellar samples with reliable membership determinations.

Photometric metallicity techniques offer a powerful complement to spectroscopy, reaching fainter magnitudes and larger spatial scales while providing metallicity estimates for statistically significant stellar samples.
Surveys such as \textit{Pristine} (\citealp{Starkenburg2017, Martin2024}, but see also \citealp{Chiti2020,Chiti2021,Placco2025}) combine a narrow-band filter centred on the Ca~H\&K absorption lines with broad-band optical photometry to estimate photometric metallicities that extend into the extremely metal-poor regime (EMP, i.e. ${\rm [Fe/H]}<-3$~dex, see \citealp{Youakim2017,Aguado2019,Venn2020,Kielty2021}).
The dedicated \textit{Pristine} dwarf galaxy survey has extended the photometric approach to a set of MW faint satellites \citep{Longeard2018,Longeard2020,Longeard2021}, already enabling the identification of their most distant and metal-poor member stars \citep{Longeard2022,Longeard2023,Longeard2025}. The benefits of this approach has also been confirmed by other investigations based on narrow-band filters \citep[e.g.][]{Fu2023,Hong2025,Pan2025}.

Our main objective in this work is to establish the metallicity distribution of MW satellites with $10^2 \lesssim L_V/L_\odot \lesssim 10^5$,  to reveal the possible presence of EMP candidates and to assess the membership of metal-rich stars. 
To this end, we revisited the initial \textit{Pristine} photometric calibration \citep[][]{Starkenburg2017}, adapting it to the broadband filters we required and extending the calibration to the cooler temperatures and redder colours typical of dwarf galaxies.
This revision allows us to sample dwarf galaxies along the red giant branch, down to the main sequence for the nearest systems, reaching $g \lesssim 23$ mag, on unprecedented spatial scales ($5$–$8 \times R_h$), more than doubling the number of known stellar members to date \citep[e.g.] []{Battaglia2022}. 

This paper is structured as follows: In Sect.~\ref{sec:data}, we describe the photometric datasets used for our sample of MW satellites. Section~\ref{sec:methods} is dedicated to  determining photometric metallicities and developing a probabilistic method to remove contamination from foreground stars.
Section~\ref{sec:results} presents the resulting metallicity catalogue of probable member stars sampled from the red giant branch and main sequence of our systems. Section~\ref{sec:discussion} discusses the derived metallicity distributions, their statistical properties, and a comparison with the literature. Finally, Sect.~\ref{sec:conclusions} contains the summary and conclusions.

%%%%%%%%%%%%%%%%%%%%%%%%%%%%%%%%%%%%%%%%%%%%%%%%%%%%%%%%%%%%%%
\section{Sample and photometric datasets}
\label{sec:data}

\begin{table*}
\caption{Sample of the Milky Way satellites analysed in this work. Columns, from left to right, indicate the dwarf galaxy name, central coordinates, total V-band magnitude, distance, half-light radius along the projected major axis, and radial velocity and metallicity sources. All values from the compilation of \citet{Battaglia2022}, except for the listed magnitudes taken from the updated catalogue (January~2021) of \citet{McConnachie2012}. References in the last two columns are: (1) \citet{Geha2026}; (2) \citet{Koposov2026}; (3) \citet{Walker2023}; (4) \citet{Simon+Geha2007}; (5) \citet{Martin2007}; (6) \citet{Spencer2018}; (7) \citet{Walker2015}; (8) \citet{Kleyna2002}; (9) \citet{Longeard2023}; (10) \citet{Fu2019}; (11) \citet{Deason2012}; (12) \citet{Aden2009}; (13) \citet{Jenkins2021}; (14) \citet{Walker2009}; (15) \citet{Simon2011}; (16) \citet{Pace2020}; (17) \citet{Willman2011}; (18) \citet{Kirby2013}; (19) \citet{Francois2016}; (20) \citet{Sitnova2021}; (21) \citet{Cohen+Huang2009}; (22) \citet{Tsujimoto2017}; (23) \citet{Ou2024}; (24) \citet{Gregory2020}; (25) \citet{Spite2018}; (26) \citet{Frebel2010}; (27) \citet{Cohen+Huang2010}.} 
\label{tab:sample}
\centering
\begin{tabular}{lccccccc}
\hline
\hline
  \multicolumn{1}{c}{Galaxy} &
  \multicolumn{1}{c}{R.A.} &
  \multicolumn{1}{c}{Dec} &
  \multicolumn{1}{c}{$M_V$} &
  \multicolumn{1}{c}{$D_\odot$} &
  \multicolumn{1}{c}{$R_{h}$} &
  \multicolumn{2}{c}{References} \\
  \multicolumn{1}{c}{} &
  \multicolumn{2}{c}{[deg]} &
  \multicolumn{1}{c}{[mag]} &
  \multicolumn{1}{c}{[kpc]} &
  \multicolumn{1}{c}{[arcmin]} &
  \multicolumn{1}{c}{RV} &
  \multicolumn{1}{c}{[Fe/H]} \\
  \hline 
  Canes Venatici~I (CVnI) & 202.0091 & 33.5521   & $-8.5\pm0.2$ & $211\pm5 $ & $7.5 \pm0.2$ & 1--5        &  1--3,5,18,19  \\
  Coma Berenices (CBer)    & 186.7458 & 23.9069   & $-4.3\pm0.3$ & $42 \pm2 $ & $5.7 \pm0.3$ & 1,2,4       &  1,2,18,20     \\
  Draco (Dra)              & 260.0684 & 57.9185   & $-8.8\pm0.1$ & $81 \pm3 $ & $9.6 \pm0.1$ & 1--3,6--8   &  1--3,18,21,22 \\
  Hercules (Her)           & 247.7722 & 12.7852   & $-5.9\pm0.2$ & $137\pm11$ & $5.8 \pm0.6$ & 1,2,4,9--12 &  1,2,18,19,23,24 \\
  Leo~IV                   & 173.2405 & $-$0.5453 & $-5.0\pm0.3$ & $154\pm5 $ & $2.6 \pm0.3$ & 1--4,13     &  1--3,13,18,19 \\
  Leo~V                    & 172.7857 & 2.2194    & $-4.4\pm0.4$ & $178\pm7 $ & $1.1 \pm0.4$ & 1--3,13,14  &  1--3,13 \\
  Pisces~II (PscII)        & 344.6345 & 5.9526    & $-4.2\pm0.4$ & $183\pm14$ & $1.2 \pm0.2$ & 1--3        &  1--3,25 \\
  Segue~1 (Seg1)           & 151.7504 & 16.0756   & $-1.3\pm0.8$ & $23 \pm2 $ & $3.9 \pm0.4$ & 1,3,15      &  3,20 \\
  Ursa Major~I (UMaI)      & 158.7706 & 51.9479   & $-5.1\pm0.4$ & $97 \pm6 $ & $8.1 \pm0.3$ & 1--5        &  1--3,18 \\
  Ursa Major~II (UMaII)    & 132.8726 & 63.1335   & $-4.5\pm0.3$ & $35 \pm3 $ & $13.9\pm0.4$ & 1--5        &  1--3,18,26 \\
  Ursa Minor (UMi)         & 227.2420 & 67.2221   & $-9.0\pm0.1$ & $76 \pm4 $ & $18.2\pm0.1$ & 1,3,6,16    &  1,3,18,27 \\
  Willman~1 (Wil1)         & 162.3436 & 51.0501   & $-2.5\pm0.8$ & $38 \pm7 $ & $2.5 \pm0.2$ & 1,2,5,17    &  1,2,5\\
\hline
\end{tabular}
\end{table*}

The MW satellites analysed in this work are listed in Table~\ref{tab:sample}, together with their central coordinates and adopted literature parameters (i.e. luminosity, distance and half-light radius).
These consist of 12 systems visible from the northern hemisphere that sample the faint-end of the MW luminosity function.

\subsection{The Pristine photometric data}
For this study, we used the final internal data release from the \textit{Pristine} dwarf galaxy survey (hereafter PDGS). These data were collected using the CaHK narrow-band filter of the MegaCam imager \citep{Boulade2003}, mounted on the Canada–France–Hawaii Telescope (CFHT). 
For each dwarf galaxy, a single $\sim1\times1$ deg$^2$ MegaCam pointing was placed on the dwarf galaxy and exposed for a total of 3\,440~s, split into 4 sub-exposures. Sub-exposures were repeated when the image quality was worse than $0.8^"$, but all sub-exposures were folded into the analysis. Individual exposures were retrieved from the CFHT archive, pre-processed by \texttt{Elixir} \citep{Magnier04} to remove instrumental signatures. Each image was processed following the procedure detailed in \citet{Martin2024}. As they explained, the resulting single-exposure photometric catalogues were calibrated using \texttt{PhotCalib} and then merged through a weighted average to yield higher signal-to-noise photometry on each detected source.
This resulted in deep photometric data with uncertainties of $\delta_{CaHK} = 0.1$ at $CaHK\simeq23$.

\subsection{The broad-band photometric data}
We relied on the broad-band photometry from \citet[][hereafter M18]{Munoz2018}, who conducted a systematic deep survey of MW satellites in the outer halo. In particular, we used the publicly available M18 primary sample, consisting of 44 observed systems.

M18 data were collected with the CFHT/MegaCam and Magellan-Clay/MegaCam wide-field imagers in the \textit{g} and \textit{r} bands, with uncertainties of $\delta_{m} = 0.1$ typically at $m\simeq25$ (or $\delta_{m} = 0.01$ at $m\simeq23$). A Gaia-DR1 astrometric solution (Gaia Collaboration et al., 2016) and a photometric calibration in the SDSS-DR7 systems \citep{Abazajian2009} was applied to the data. For further details on the data reduction and photometric calibration we refer the reader to M18.  

Due to saturation, sources with $g\lesssim18$ are missing from the M18 catalogue. Furthermore, the footprint around each system did not always match that of PDGS due to differences in the mosaic area. 
Since potential members of some of our systems can be found up to $g\sim16$ and outside the M18 footprint, we complemented the M18 photometry with that of Pan-STARRS1 \citep[][hereafter PS1]{Chambers2016}. 

The PS1 catalogue is shallower than M18, reaching $\delta_{g} = 0.1$ typically at $g\simeq22$, but extends to brighter magnitudes (up to 14). Although the M18 catalogue was calibrated using SDSS, we chose PS1 because it uniformly covers the northern hemisphere and because the photometric transformation between the two systems is mostly linear in the \textit{g} and \textit{r} bands.

We used the PS1-DR1 catalogue that relies on the PSF photometry from the MeanObject table, which provides the smallest uncertainties.\footnote{For further details, we refer to: \url{https://outerspace.stsci.edu/display/PANSTARRS/PS1+Comparison+of+different+photometric+measures}.} 
This catalogue was used to calibrate each dwarf galaxy field in M18 relative to PS1 in order to ensure homogeneous photometry. The calibration procedure is reported in Appendix~\ref{apx:sample}. The residuals between the calibrated and reference PS1 magnitudes typically averaged around zero, with a small standard deviation of $0.02-0.03$~mag.  

Unless stated otherwise, magnitudes and colours reported in the rest of the text are corrected for extinction. We used the \citet{S+F2011} recalibration of the \citet*{SFD1998} dust maps, employing the extinction coefficients provided in \citet{S+F2011} and \citet{Starkenburg2017} for the CaHK band.

\subsection{Cross-matching and cleaning}
We matched the PDGS catalogue with M18 and PS1 separately, keeping them distinct in the analysis until the final metallicity sample was created (see Sect.~\ref{sec:results}). 
We therefore selected sources that are consistent with stellar detections using the morphological indices (i.e. $\verb!chi!$ and $\verb!sharp!$) provided by M18, applying a star-galaxy separation for PS1 which uses the difference between PSF and Kron magnitudes \citep[e.g.][]{Farrow2014}.

Specifically, for M18 we kept targets with $\verb!chi!<3$ and $ \left | \verb!sharp! \right |<0.5$, whilst for PS1 we kept targets with $g_{\rm PSF}-g_{\rm Kron} \leq 0.05$ and $r_{\rm PSF}-r_{\rm Kron} \leq 0.05$.\footnote{See \url{https://outerspace.stsci.edu/display/PANSTARRS/How+to+separate+stars+and+galaxies}}
We then selected sources with $\delta_{CaHK}\leq0.1$, applying the same cut-off to \textit{g} and \textit{r} for PS1, or reducing it to 0.01 for M18. This ensured a balance between photometric quality and number of detections, resulting in clean catalogues with $28\,321$ sources for M18 and of $40\,465$ for PS1. 

The final step in preparing the catalogues was to include, where available, \textit{Gaia} information and literature spectroscopic data for each source. For the former, we relied on the third \textit{Gaia} data-release \citep[DR3,][]{GaiaDR3}, including information for detected objects with a full and high quality astrometric solution ($\verb!astrometric_params_solved!\geq31$,  $\verb!ruwe!<1.4$ and $\verb!astrometric_excess_noise_sig!< 2$) and that are not flagged as duplicates ($\verb!duplicated_source! = \verb!False!$).

Additionally, we included a variability flag by cross-matching with \citet{Gavras2023}, who provided a catalogue of known variable objects from the literature present in Gaia-DR3. This includes both stellar and extragalactic variable sources. We removed sources marked as RR-Lyrae and Cepheids (likely belonging to our systems), as well as the AGNs, including the QSOs used for the \textit{Gaia} celestial reference frame \citep[CRF3,][]{Gaia-CRF3}. This reduced the number of targets by around 2\%.

We have incorporated several spectroscopic studies that report radial velocities and metallicities for our systems (see Table~\ref{tab:sample}). As described in Sect.~\ref{subsec:membership}, radial velocities were included in the likelihood terms used to identify member stars in our systems. In contrast, metallicities were used mainly to validate our method and compare with our derived results.
To reduce biases when merging the different tables, we adopted a homogeneous approach, namely prioritising publications that analysed multiple systems at once \citep[mainly][]{Walker2023,Geha2026,Koposov2026}.

We note that values from the DESI-DR1 catalogue \citep{Koposov2026} were included after applying the following cuts to select likely stellar sources: $\verb!RVS_WARN!=0$, $\verb!RR_SPECTYPE!=``{\rm STAR}"$, $\verb!PRIMARY!={\rm True}$, and $\verb!VSIN!<30$ for radial velocities generated with the RVS pipeline, while $\texttt{BESTGRID} \ne \texttt{`s\_rdesi1'}$ and $\texttt{SUCCESS}=1$ for metallicities generated with the SP pipeline. These were then recalibrated accordingly \citep[see again][their Sect.~4.2.1]{Koposov2026}. 

We further retained radial velocity measurements with $S/N_{\rm RV}>2$, within a range of $\pm500$~km\,s$^{-1}$, and with uncertainties $<50$~km\,s$^{-1}$. This rather broad selection was intended to facilitate the identification of foreground contaminants while including as many values as possible. For the metallicity measurements, we retained sources with $S/N_{\rm [Fe/H]}>10$ and uncertainties $<0.2$~dex.

%%%%%%%%%%%%%%%%%%%%%%%%%%%%%%%%%%%%%%%%%%%%%%%%%%%%%%%%%%%%%%
\section{Methods}
\label{sec:methods}

The determination of photometric metallicity catalogues for our sample of dwarf galaxies required two main steps. The first was to obtain the membership probability of each star in a given system. The second was to compute the photometric metallicity using the \textit{CaHK} magnitudes together with the associated broad-band photometry.

\subsection{Membership determination}
\label{subsec:membership}

Due to the low surface brightness of our systems, most sources detected in a given field will be MW contaminants. Therefore, we needed a flexible method that could incorporate the available photometric and spectroscopic information on each star to determine its membership probability to a given system.

We applied an updated version of the maximum likelihood method presented in \citet{Battaglia2022}, which is based on the work of \citet{McConnachie+Venn2020} and \citet{Pace+Li2019}. We refer the reader to these sources for a detailed description of the methodology. Here, we report the essential, highlighting the changes we have introduced.

We adopted a mixture model in which each star was considered either to belong to the dwarf galaxy being studied or to be a contaminant. The total likelihood $\mathcal{L}$ is then:
\begin{equation}
    \mathcal{L} = f_{\rm sat}\mathcal{L}_{\rm sat} + (1-f_{\rm sat})\mathcal{L}_{\rm MW}
\end{equation}
where $\mathcal{L}_{\rm sat}$ and $\mathcal{L}_{\rm c}$ are the respective likelihood terms for the satellite and the MW contamination, and $f_{\rm sat}$ is the fraction of satellite member stars. Each likelihood is decomposed into several terms that account for the available information on each star:
\begin{equation}
    \mathcal{L}_{\rm sat/MW} = \mathcal{L}_{\rm Sp}\mathcal{L}_{\rm CMD}\mathcal{L}_{\rm PM}\mathcal{L}_{\rm RV}
\end{equation}
where $\mathcal{L}_{\rm Sp}$, $\mathcal{L}_{\rm CMD}$, $\mathcal{L}_{\rm PM}$ and $\mathcal{L}_{\rm RV}$ are the terms for spatial distribution, position on the colour–magnitude diagram, proper motion and radial velocity information, respectively. 

We maximised the total likelihood function within a Bayesian framework, where the only free parameter was $f_{\rm sat}$, since we were only interested in determining the probability of each star belonging to a given satellite. We assumed a uniform prior for the fraction parameter ranging between $0<f_{\rm sat}<1$. We used the \texttt{MultiNest} algorithm \citep{Feroz+Hobson2008,Feroz2009} to determine the posterior distribution.

Finally, membership probabilities were calculated as follows:
\begin{equation}
    \mathcal{P} = \frac{f_{\rm sat}\mathcal{L}_{\rm sat}}{f_{\rm sat}\mathcal{L}_{\rm sat} + (1-f_{\rm sat})\mathcal{L}_{\rm MW}}
\end{equation}
where we denote as $\mathcal{P}_{\rm i}$ the individual star's probability. 

The individual likelihood terms were in general determined as in \citet{Battaglia2022}, either as empirical or analytical 2D look-up maps. However, we made a few changes due to the differences introduced by our data. We refer to Appendix \ref{apx:likelihood} for a detailed description of the likelihood and the changes introduced.

\subsection{Metallicity determination}
\label{subsec:photo-FeH}

\begin{figure*}
    \centering
    \includegraphics[width=\textwidth]{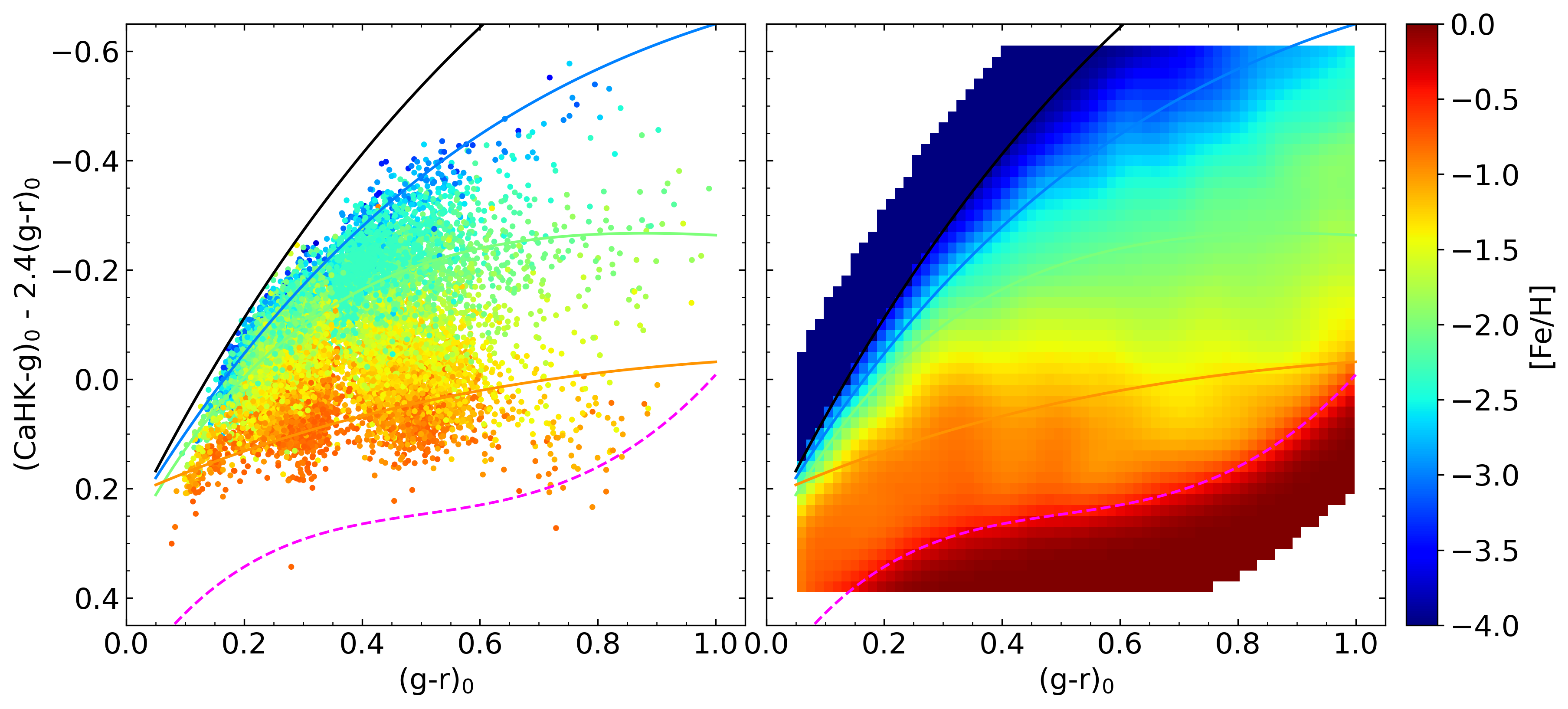}
    \caption{Pristine colour-colour diagram with the calibration sample distribution (\textit{left}) and the estimated calibration grid (\textit{right}). The colour-map indicates the spectroscopic metallicity of each star. Iso-metallicity lines for values of ${\rm [Fe/H]} = [0,-1,-2,-3]$~dex, as well as no metals, are shown respectively from bottom to top.}
    \label{fig:grid}
\end{figure*}

We obtained photometric metallicity values ${\rm [Fe/H]_{Pr}}$ for individual stars in each dwarf galaxy, using the method presented in \citet{Starkenburg2017} and \citet{Martin2024}. Briefly, this method uses a calibration sample to create a mapping between a colour-colour space (which combines \textit{CaHK} and broadband photometry) and [Fe/H] values. In practice, each stellar value on the colour-colour diagram is interpolated with a [Fe/H] grid generated from the calibration sample. For this, we used the algorithm made available in \citet{Martin2024}, with the necessary modifications.\footnote{See \url{https://github.com/ankearentsen/Pristine_code}}

The main difference between our analysis and previous \textit{Pristine} studies is the use of a colour-colour space that (necessarily) includes the $r$ band instead of the $i$ band \citep[as in e.g.][]{Starkenburg2017,Longeard2018}. The $(g-r)_0$ colour is temperature-sensitive \citep{Ivezic2008} and can be combined in the same way with the \textit{CaHK} band to remove its temperature dependence whilst retaining its dependence on metallicity \citep{Starkenburg2017}. As shown in Fig.~\ref{fig:grid}, this led to a colour-colour space defined as: $(CaHK-g)_0-2.4\times(g-r)_0$ vs $(g-r)_0$, where the synthetic iso-metallicity lines are adapted from \citet{Starkenburg2017} by mean of colour transformation.

The calibration sample shown in the figure includes spectroscopic ${\rm [Fe/H]_{Sp}}$ measurements for MW stars obtained from various surveys that fall within the \textit{Pristine}-DR1 footprint \citep[see][and references therein]{Martin2024}. 
From these sources we selected giants with $T_{\rm eff}<5800$~K and ${\rm log}(g)<3.8$~dex, as well as main sequence stars, with $T_{\rm eff}>5800$~K, within the colour range covered by our dwarf galaxies (i.e. $0.0<(g-r)_0<1.0$). This selection corresponds to the locations of giant and main sequence stars in our dwarf galaxies on the colour-colour plane. These two populations separate at $(g-r)_0\sim0.3$, thus enabling us to apply a single calibration (see Fig.~\ref{fig:grid}).

After cross-matching with PS1, we restricted the calibration sample to stars with photometric uncertainties $<0.05$~mag and ${\rm [Fe/H]_{Sp}}$ uncertainties $<0.25$~dex. This also resulted in the exclusion of APOGEE stars due to their poor broadband photometry.\footnote{We note that the majority of APOGEE stars had PS1 magnitudes close to the saturation limit ($g,r\lesssim14$~mag; \citealp{Magnier2013}).} 
To improve sampling in the metal-poor regime (i.e.~${\rm [Fe/H]_{Sp}}<-2$~dex), we also included stars with observed \textit{CaHK} from globular clusters and dwarf galaxies. For the former, we included the metal-poor systems NGC~5024, NGC~6341 and NGC~7078, assigning them the cluster average metallicity \citep{Carretta2009}. They were selected within the \textit{Pristine}-DR1 footprint, according to the criteria outlined in \citet{Martin2024}, and then cross-matched with PS1.

For the latter, we used the compilation of dwarf galaxy stellar metallicities from the SAGA database \citep{Suda2017}, selecting values obtained at medium-to-high resolution.
We also included stellar metallicities from the dwarf galaxy high-resolution compilation of \citet{Walker2023}. In both cases, ${\rm [Fe/H]_{Sp}}$ values were included with an uncertainty $<0.20$~dex. Their \textit{CaHK} comes from our Pristine-DG catalogues instead. This added a further 1207 stars to the sample, bringing the total to $\sim 19\,500$. Of these, 95\% had uncertainties $<0.01$~mag and $<0.10$~dex, respectively in magnitude and ${\rm [Fe/H]_{Sp}}$.

The colour-colour space was then divided into a grid limited to the range $0.0<(g-r)_0<1.0$ and $-0.6<(CaHK-g)_0-2.4\times(g-r)_0<0.4$, with the 2$\sigma$-clipping average metallicity of the calibration sample assigned to each cell. 
The grid was smoothed by applying a Gaussian kernel and extended along the y-axis, setting cells to ${\rm [Fe/H]} =-4.0$~dex up to 0.2~mag below the “metal-free” line and to ${\rm [Fe/H]} =0.0$~dex up to 0.2~mag above the “solar metallicity” line. The latter line was obtained by fitting a third degree polynomial to those stars with ${\rm [Fe/H]_{Sp}}$ within 0.15~dex from the solar value.

We note that the metallicity separation on the calibration plane is greatly reduced for stars with ${\rm [Fe/H]_{Sp}}\gtrsim-1.0$~dex, mainly due to the increased impact of gravity on the CaHK lines \citep{Starkenburg2017}. Therefore, within the limits represented by the “metal-free” and “solar metallicity” lines, we restricted the calibration sample to values of ${\rm [Fe/H]_{Sp}}<-0.75$~dex to remove this bias  (see again Fig.~\ref{fig:grid}).

Finally, metallicities were obtained as posterior distributions by interpolating the observed values over the calibration grid through a $10^4$ Monte Carlo sampling of the magnitude uncertainties.
For each star, we stored the ${\rm [Fe/H]_{Pr}}$ value corresponding directly to the observed position on the colour-colour grid, together with the 16th, 50th and 84th percentiles of the posterior distribution (hereafter simply ${\rm [Fe/H]_{Pr}}$ and ${\rm [Fe/H]_{16th}},{\rm [Fe/H]_{50th}},{\rm [Fe/H]_{84th}}$). 
We also stored the fraction of the Monte Carlo samples \texttt{mcfrac} that fell within the limits of the metallicity grid. In the following, we approximate the photometric metallicity uncertainty as $\delta{\rm [Fe/H]_{Pr}} = ({\rm [Fe/H]_{84th}} - {\rm [Fe/H]_{16th}})/2$.

%%%%%%%%%%%%%%%%%%%%%%%%%%%%%%%%%%%%%%%%%%%%%%%%%%%%%%%%%%%%%%
\section{Results}
\label{sec:results}

The M18 and PS1 catalogues, presented in Section~\ref{sec:data}, were analysed using the method described in Section~\ref{sec:methods} to obtain membership probabilities and photometric metallicities for each star in both catalogues. 
These data were compiled into a single catalogue, validated, and used to construct the metallicity distributions of individual systems.

\subsection{The photometric metallicity catalogue}
\label{subsec:photo-met-cat}

We merged the M18 and PS1 catalogues, selecting all the stars in PS1 that were not in M18, or that had $g_{\rm M18}\leq18.5$. To match the higher photometric quality of the M18 catalogue, we only retained stars from PS1 with photometric uncertainties of at most 0.01~mag in \textit{g} and \textit{r}, unless they had membership probability $\mathcal{P}_{\rm i}\geq0.05$, in which case we relaxed this condition to 0.05~mag. The aim was to avoid excluding potential members that might be in the transition zone between the two catalogues, where PS1 photometry becomes less accurate. This resulted in a final merged catalogue of $38\,824$ stellar sources.\footnote{We note that we did not perform any completeness tests on our final catalogue, given its cross-matched nature and the fact that a selection based on magnitude uncertainties effectively define a magnitude-dependent completeness limit. However, probable members were found up to $g_0\sim23$, which is well above the limiting magnitudes of the M18 and CaHK catalogues.}

Since we did not apply any preliminary cut that could remove the foreground contaminants, only about 15\% of the stars in the entire catalogue have $\mathcal{P}_{\rm i}\geq0.05$ of belonging to a dwarf galaxy. This is not surprising given the low surface brightness of our systems. Nevertheless, the high photometric quality of our dataset has resulted in average metallicity uncertainties of around 0.2~dex.

We performed a series of internal consistency and external validation tests to further assess the quality of the merged catalogue and calculated photometric metallicities. Internal consistency was verified by calculating photometric metallicities for the calibration sample and comparing these with spectroscopic values. Additionally, we compared the photometric metallicities of stars that appeared in both the PS1 and M18 catalogues. The external validation was performed by cross-matching the merged catalogue with a compilation of high-resolution spectroscopic studies and the spectroscopic survey data from \citet{Geha2026}, DESI-DR1 \citep{Koposov2026} and \citet{Walker2023}.

As detailed in Appendix~\ref{apx:results}, we found no significant internal biases that could have been introduced during the estimation of photometric metallicities or the merging of the M18 and PS1 catalogues. The external tests also revealed no significant biases in the residuals (except on a system-by-system basis with the \citealp{Geha2026} compilation), with a limited spread ($\sim0.20$~dex). This shows very good agreement with the spectroscopic literature and reinforces the validity of our results.

However, care must be taken when interpreting the metallicity values obtained for the faintest stars. Given the complex mapping of photometric values to metallicity ones, photometric dispersion at low magnitudes can skew metallicity estimates towards the extremes of the calibration grid \citep[see also][]{Martin2024}. Although the quality cuts described in the following section greatly limits this effect, for those systems where the main sequence is sampled (ComBer, UMaII, Seg1, Wil1), the metallicity values may still be biased towards higher values due to the reduced metallicity separation on the calibration diagram for $(g-r)_0<0.4$ (see Fig.~\ref{fig:grid}).

We further note that the calculated ${\rm [Fe/H]_{Pr}}$ of stars with strong carbon absorption may be significantly overestimated, as shown in \citet{Martin2024}. This considerably limits our ability to identify carbon-enhanced metal-poor (CEMP) stars; that is stars with ${\rm [C/Fe]}>0.7$~dex and ${\rm [Fe/H]}<-2.0$~dex \citep[e.g.][]{Spite2013}. As shown in Appendix~\ref{apx:results}, we have recovered ${\rm [Fe/H]_{Pr}}$ values for known CEMPs that were up to 1~dex higher than those reported in the literature. Below, we have simply excluded from our analysis the known CEMPs that appear in our sample, leaving the task of identifying any new ones to future studies (e.g. Montelius in prep.).

\subsection{Metallicity updated membership probabilities}
\label{subsec:Pf}

In order to build meaningful metallicity distributions and limit the contribution of foreground stars, we needed to refine the membership estimation by including the obtained metallicity information. To keep measurements within the limits of the calibration grid, we applied first the following selections: ${\rm [Fe/H]_{16th}<0.0}$~dex, ${\rm [Fe/H]_{84th}>-4.0}$~dex, and $\texttt{mcfrac}>0.8$. 

At the same time we kept uncertainties within $\delta{\rm [Fe/H]_{Pr}}<0.5$~dex and $\delta_{CaHK}<0.05$~mag \citep[see also][their Sect.7.3]{Martin2024}. We note that we did not apply the latter condition yo the faintest systems: Leo~IV, Leo~V, Pisces~II, Ursa~Major~I, and Willman~1, in order to avoid being too restrictive given their low member statistics ($\lesssim50$). 

The next step in deriving the metallicity distribution was to select probable member stars according to a sensible $\mathcal{P}_{\rm i}$ threshold. However, to avoid including too many high-metallicity contaminants ($\gtrsim-1$ dex), we compared the metallicity distributions of very likely members and contaminants, using this information to refine the final membership probability.

In practice, we modelled the distributions of RGB stars with $\mathcal{P}_{\rm i}\ge0.95$ and of probable contaminants with $\mathcal{P}_{\rm i}\le0.05$ separately, using a Gaussian kernel density estimate with a width of 0.25~dex. Considering them as the metallicity likelihood functions $\mathcal{L}_{\rm [Fe/H]}$ of the satellite and the contamination, and applying the Bayes theorem, the updated membership probabilities became:
\begin{equation}
    \mathcal{P}_{\rm f} = \frac{\mathcal{P}_{\rm i}\mathcal{L}_{\rm sat,\,[Fe/H]}}{\mathcal{P}_{\rm i}\mathcal{L}_{\rm sat,\,[Fe/H]} + (1-\mathcal{P}_{\rm i})\mathcal{L}_{\rm MW,\,[Fe/H]}}
\end{equation}

The benefits of updating the individual probabilities were most evident in systems where we sample down to the main sequence, such as in Coma~Berenices, Segue~1, Ursa~Major~II, and Willman~1. In these cases, the main sequence lies below both the Gaia limit ($g_0\sim21$) and the available spectroscopy. Consequently, the initial probabilities were based solely on the spatial and photometric likelihoods, with a greater chance of introducing MW contaminants as false positives. The inclusion of metallicity information reduces this possibility. 

We proceeded by selecting probable member stars with $\mathcal{P}_{\rm f}\ge0.8$ to build the metallicity distribution of each system. 
This resulted into 3\,917 selected stars, with an average $\delta{\rm [Fe/H]_{Pr}}$ of $\sim 0.15$~dex.
Compared to \citet{Battaglia2022}, we more than double (over an equivalent area) the number of recovered probable members in each system (see Table~\ref{tab:FEH-dist}), with the additional stars mostly found at fainter magnitudes than the Gaia limit.

The selected stars allowed us to calculate the average metallicity of each system (as error-weighted mean and median values), the metallicity spread (as standard deviation and median absolute deviation), as well as to estimate the fraction of extremely metal-poor (${\rm [Fe/H]}<-3.0$~dex) stars, as reported in Table~\ref{tab:FEH-dist}. 
Individual diagnostic plots showing the spatial distribution and photometric properties of the member stars, including their uncertainties, are shown in Figs.~\ref{fig:FEH-dist-UFD-1}-\ref{fig:FEH-dist-UFD-4} in the Appendix.

\begin{table*}
\caption{Properties of the metallicity distributions. Columns, from left to right, indicate the dwarf galaxy acronym, the number of stars in each field, the number of probable member stars with $\mathcal{P}_{\rm f}\ge0.05$ and $\ge0.8$, the average metallicity (as error-weighted mean and median), the metallicity scatter (as standard deviation and median absolute deviation), the skewness and excess kurtosis, the number of extremely metal-poor (${\rm [Fe/H]}<-3.0$~dex) stars.} 
\label{tab:FEH-dist}
\centering
\begin{tabular}{lrrrrrrrrrrr}
    \hline
    \hline
	  \multicolumn{1}{c}{Galaxy} &
	  \multicolumn{1}{c}{$N_i$} &
	  \multicolumn{2}{c}{$\mathcal{P}_{\rm f}$} &
	  \multicolumn{2}{c}{$\mu_{\rm [Fe/H]}$} &
	  \multicolumn{2}{c}{$\sigma_{\rm [Fe/H]}$} &
    \multicolumn{1}{c}{$\gamma_{1,\rm [Fe/H]}$} &
    \multicolumn{1}{c}{$\gamma_{2,\rm [Fe/H]}$} &
	  \multicolumn{1}{c}{$N_{\rm EMP}$} \\
	  \multicolumn{1}{c}{} &
	  \multicolumn{1}{c}{} &
	  \multicolumn{1}{c}{$\ge0.05$} &
	  \multicolumn{1}{c}{$\ge0.8$} &
	  \multicolumn{1}{c}{$M_W$} &
	  \multicolumn{1}{c}{$M_{50}$} &
	  \multicolumn{1}{c}{STD} &
	  \multicolumn{1}{c}{MAD} &
    \multicolumn{1}{c}{} &
    \multicolumn{1}{c}{} &
	  \multicolumn{1}{c}{} \\
    \hline
    CVnI   & 2763 & 417  & 330  & $-2.18$ & $-2.30$ & $0.38$ & $0.33$ & $-0.69$ & $1.28$  & 17 \\
    ComBer  & 2406 & 173  & 121  & $-2.33$ & $-2.33$ & $0.38$ & $0.43$ & $-0.27$ & $-0.48$ & 7  \\
    Dra     & 6693 & 1394 & 1339 & $-2.15$ & $-2.29$ & $0.36$ & $0.27$ & $-0.89$ & $2.25$  & 62 \\
    Her     & 9117 & 96   & 74   & $-2.38$ & $-2.58$ & $0.35$ & $0.28$ & $-0.52$ & $1.90$  & 8  \\
    LeoIV   & 1273 & 27   & 25   & $-2.30$ & $-2.26$ & $0.38$ & $0.30$ & $-0.75$ & $0.77$  & 1  \\
    LeoV    & 1099 & 14   & 14   & $-2.19$ & $-2.24$ & $0.37$ & $0.34$ & $0.55$  & $-0.41$ & 0  \\
    PscII   & 1899 & 13   & 10   & $-2.43$ & $-2.39$ & $0.74$ & $0.20$ & $-0.70$ & $-0.17$ & 2  \\
    Seg1    & 1739 & 42   & 29   & $-1.71$ & $-2.62$ & $0.65$ & $0.81$ & $0.10$  & $-1.27$ & 6  \\
    UMaI    & 1406 & 67   & 51   & $-2.42$ & $-2.60$ & $0.53$ & $0.35$ & $-0.22$ & $0.80$  & 10 \\
    UMaII   & 3214 & 273  & 185  & $-2.28$ & $-2.22$ & $0.44$ & $0.43$ & $-0.44$ & $1.05$  & 4  \\
    UMi     & 5594 & 1801 & 1710 & $-2.12$ & $-2.21$ & $0.36$ & $0.29$ & $-0.76$ & $1.98$  & 51 \\
    Wil1    & 1621 & 33   & 29   & $-2.29$ & $-2.30$ & $0.42$ & $0.48$ & $-0.15$ & $-0.72$ & 2  \\  
    \hline
    Total  & 38\,824 & 4\,350 & 3\,917 &  &  &  &  &  &  & 170 \\
    \hline
\end{tabular}
\end{table*}

%%%%%%%%%%%%%%%%%%%%%%%%%%%%%%%%%%%%%%%%%%%%%%%%%%%%%%%%%%%%%%
\section{Discussion}
\label{sec:discussion}

\subsection{Metallicity distributions}
\label{subsec:mdfs}

\begin{figure*}
    \centering
    \includegraphics[width=.99\textwidth]{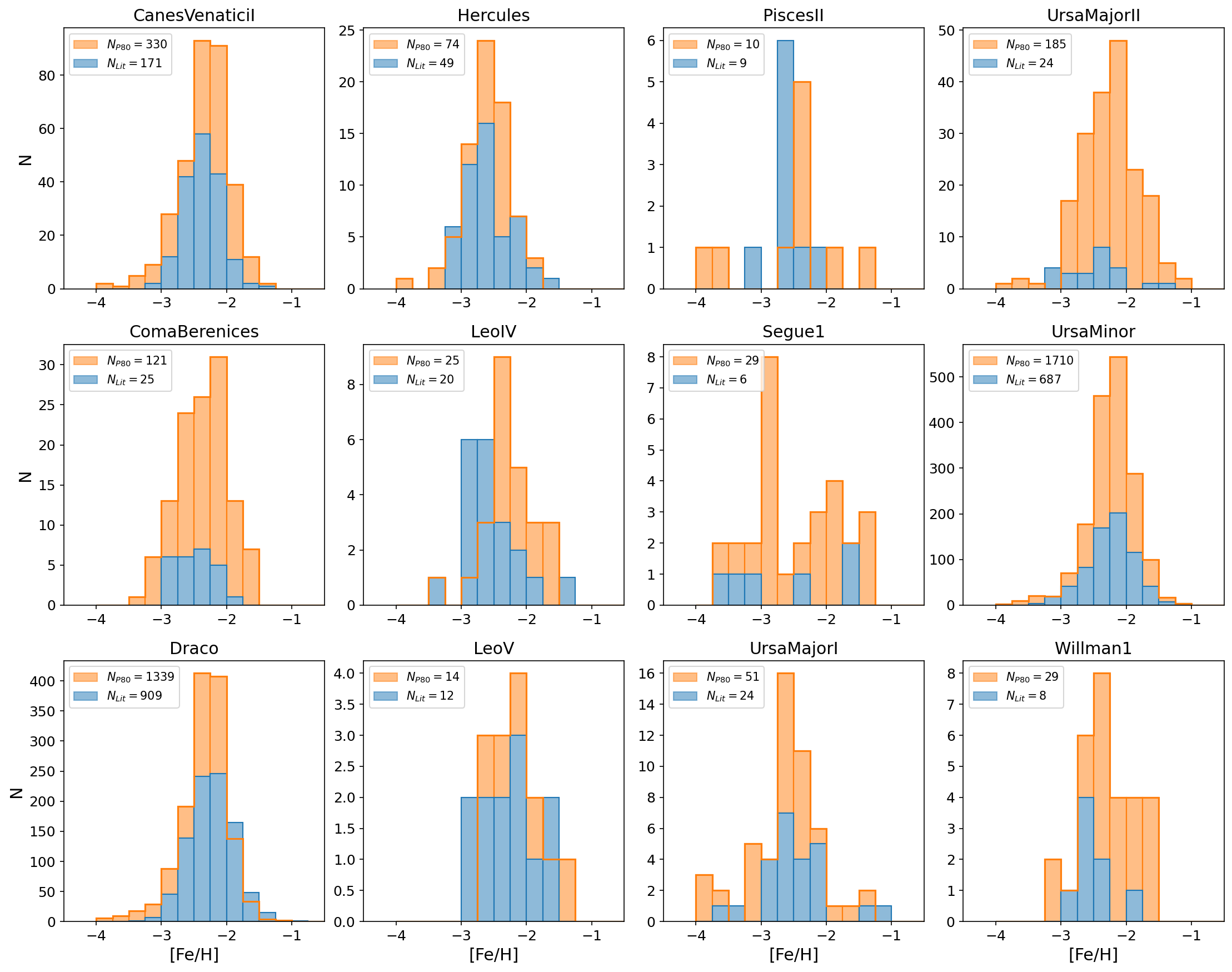}
    \includegraphics[width=.99\textwidth]{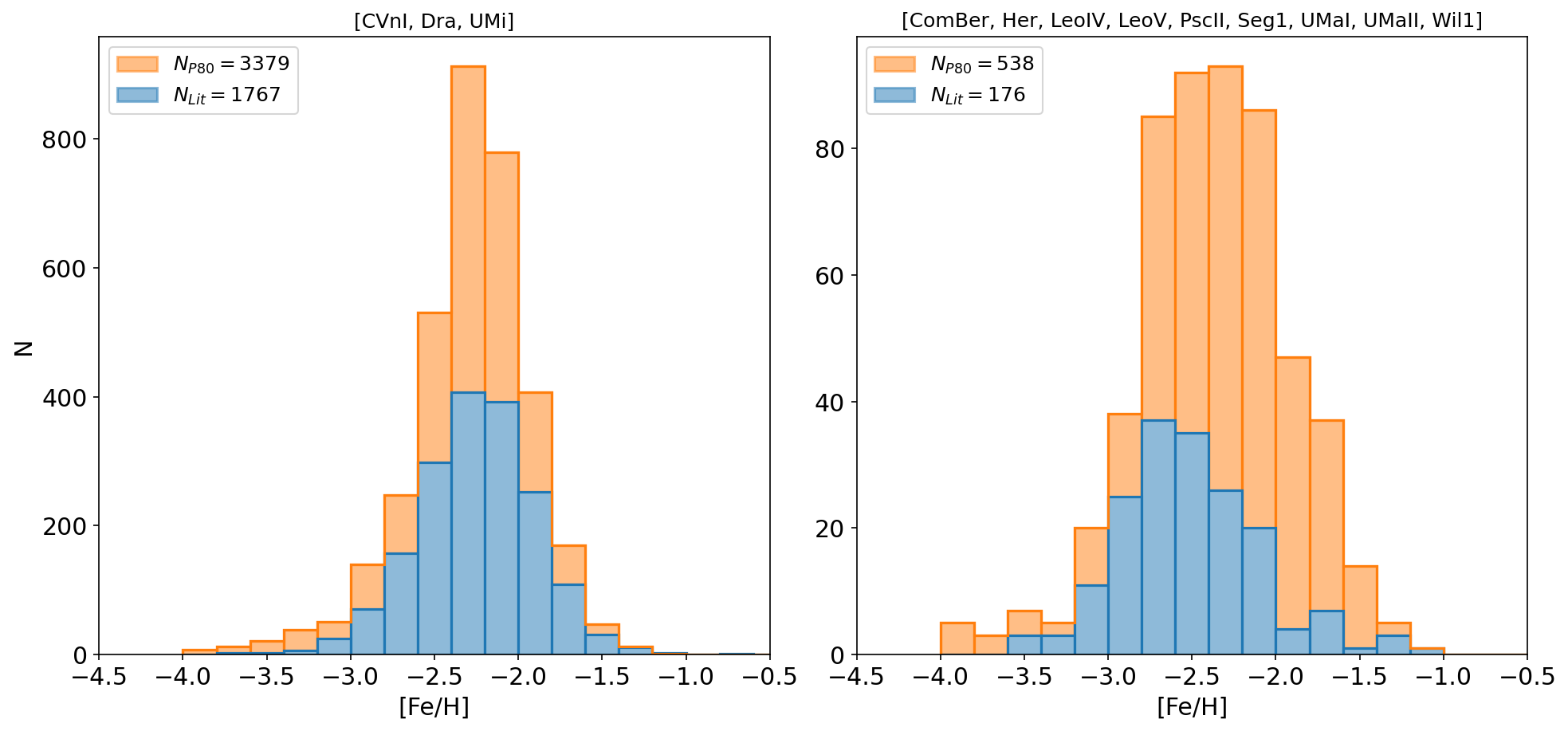}
    \caption{
    \textit{Top:} Metallicity distributions for each system studied in this work. Each panel shows the distribution of probable members (in red), alongside the distribution of members found in literature on spectroscopic work (in blue; see references in Table~\ref{tab:sample}).
    \textit{Bottom:} Combined distributions for the brighter systems (CVnI, Dra, and UMi; \textit{left}) and the other UFDs analysed in this work (\textit{right}).
    }
    \label{fig:Hist-FeH}
\end{figure*}

The metallicity distribution (MD) in a galaxy depends on its star formation history and on how this relates to the evolution of its gas content. This dynamic is influenced by accretion and outflow processes, such as those caused by stellar feedback or tidal stripping \citep{Lin+Faber1983,Dekel+Silk1986,Marcolini2006,Governato2010}. 
While the MD of the brightest MW satellites have been well characterised over time, this has not been the case for most of the UFD satellites due to the small spectroscopic samples collected \citep[see][and references therein]{Simon2019}. 

We present here an overview of the MD obtained for our sample of galaxies, alongside a dedicated discussion section for each system. We compare our results with available spectroscopy to demonstrate that photometric metallicity estimations are a valid complement for overcoming spectroscopic limitations in terms of photometric depth and target selection (see also \citealp{Longeard2018,Longeard2020,Longeard2021}; but also \citealp{Pan2025,Atzberger2026arXiv}).

Figure~\ref{fig:Hist-FeH} shows the MD of probable member stars in each galaxy, as well as the combined MDs of the dSphs (CVnI, Dra, UMi) and the UFDs (ComBer, Her, LeoIV, LeoV, PscII, Seg1, UMaI, UMaII, Wil1). The properties of each MD are summarised in Table~\ref{tab:FEH-dist}.
The dSphs show the most regular distributions, characterised by a well-defined peak, moderate dispersion ($\sigma_{\rm MAD} \sim 0.35$--$0.40$~dex), and negative skewness indicating an asymmetric tail toward lower metallicities. 
Most UFDs in our sample, although they exhibit dispersions comparable to those of dSphs, have individual MDs that are noisier and less populated, making it more difficult to meaningfully characterise their individual shapes. We refer to Sect.~\ref{subsec:glx} for a dedicated analysis of the individual MDs.

Figure~\ref{fig:Hist-FeH} also shows a comparison with the literature. Spectroscopic values were selected from the sources listed in Table~\ref{tab:sample} for common targets with $\mathcal{P}_{\rm f}\ge0.8$, $\delta{\rm [Fe/H]}_{\rm Sp}\leq0.5$~dex and $S/N>5$.  
Significant improvements over the existing knowledge are evident for each of the individual galaxies systems and for their combination (see the bottom panels of Fig.~\ref{fig:Hist-FeH}). In the ultra-faint regime in particular, the number of probable member stars has increased by more than a factor of two, thanks to our ability to reach fainter magnitudes and to examine the outermost regions. 

When we normalise the combined photometric and spectroscopic MDs displayed in the lower panels of Fig.~\ref{fig:Hist-FeH}, the dSphs show overlapping distributions. The UFDs also present broadly similar distributions, but their mean values differ by about $\sim 0.1$~dex, with most photometric [Fe/H] measurements lying at ${\rm [Fe/H]}>-2.5$~dex. This offset arises because our spectroscopic compilation is largely based on data from \citet{Geha2026}, for which we identify a systematic bias of $\sim 0.1$~dex when we compare stars in common for the faint systems (see Appendix~\ref{apx:results}). Accounting for this bias would reconcile the two histograms.

\subsubsection{Extremely metal-poor and notable metal-rich stars}
\label{subsec:emp-mr}

The improved sampling of our photometric catalogue allowed us to identify stars at the extreme ends of the metallicity distribution in several systems for the first time. As shown in Table~\ref{tab:FEH-dist}, we found a total of 170 candidate extremely metal-poor (EMP) stars with ${\rm [Fe/H]} < -3.0$~dex. We also identified a number of metal-rich (MR) stars in each system with ${\rm [Fe/H]} \gtrsim -1.5$~dex that are detached from the rest of their respective distributions. 

Candidate EMP stars are found in nearly all of our systems. Most are in the best-surveyed systems, but their occurrence in UFDs is notable given the rarity of such stars in current spectroscopic samples (see Table~\ref{tab:sample}). In general, we find EMP candidates at all magnitudes, although their fraction tends to increase at the faint magnitude limit. As discussed in Sect.~\ref{subsec:photo-met-cat}, photometric uncertainties could disperse the metallicity values towards the extremes of the calibration grid. Thus, while purity is expected to decrease at fainter magnitudes, the key point is that we detect EMP candidates at all magnitudes. A complete spectroscopic follow-up would firmly establish the EMP distribution in UFDs and enable a more robust comparison with models \citep[see also discussion in][]{Sanati2023}.

At the MR end of the distributions, several stars require cautious interpretation. As discussed, we expect to find MR stars with large uncertainties, especially in systems where the main sequence is sampled (ComBer, UMaII, Seg1, Wil1). However, we also find about ten bright RGB stars in individual systems that are much more metal-rich than the rest of their sample. Below, we examine them individually, noting their dubious membership and likely foreground origin.

\subsubsection{Comments on individual systems}
\label{subsec:glx}

\paragraph{\textbf{Canes Venatici I.}}
Canes~Venatici~I has a similar luminosity to Draco and Ursa~Minor, but is located at a greater heliocentric distance ($\sim200$~kpc; see Table~\ref{tab:sample}). This has allowed us to cover most of the galaxy within a single pointing, as shown in Fig.~\ref{fig:FEH-dist-UFD-1}. 

The 330 probable members sample the RGB of the galaxy down to its horizontal branch. Their spatial distribution follows the ellipticity of the galaxy and are detected out to a maximum projected distance of $\sim 6 \times R_h$. The MD peaks at ${\rm [Fe/H]} \simeq -2.3$~dex, has a scatter comparable to the other bright systems ($\sigma_{\rm [Fe/H]} \sim 0.35$~dex), and shows a declining radial profile. 

Our results represent a significant improvement on the existing spectroscopic literature (see Table~\ref{tab:sample}), both in terms of size (by almost a factor of two) and spatial extension (${\rm [Fe/H]}_{\rm Sp}$ values are mostly within $2\times R_h$).
We clearly uncover a metallicity gradient (see Fig.~\ref{fig:FEH-dist-UFD-1}) that was only hinted at in previous studies \citep[e.g.][]{Kirby2011,Han2020}. A least-square linear fit gives a value of $\sim-0.1$~dex\,$R_h^{-1}$, within the distribution of Local Group dwarf galaxies \citep{Taibi2022}.

We identify 17 EMP candidates, most of which have no spectroscopic information available. At the MR end, two stars detach from the rest, with values on the CaHK calibration plane close to the contaminants sequence (see Fig.~\ref{fig:FEH-dist-UFD-1} again). However, when taking their metallicity uncertainties into account, they are still within the MD, as for their corresponding spectroscopic values \citep[][]{Geha2026b}.

We finally note that our analysis does not include the CEMP-no star identified by \citet{Yoon2020}. Its very red colour, $(g-r)_0\gg 1.0$, places it outside our photometric selection window.

\paragraph{\textbf{Coma~Berenices.}}
Coma~Berenices hosts a stellar population that is old ($>12$~Gyr) and metal-poor \citep{Brown2014, Durbin2025}, which we cover in its entirety within the survey’s coverage area, as shown in Fig.~\ref{fig:FEH-dist-UFD-1} (central panel). 
We found 121 probable members extending down to the main sequence and reaching as far as $5\times R_h$. 

Metallicities show a negatively skewed distribution, peaking at ${\rm [Fe/H]} \simeq -2.3$~dex, with a moderate scatter $\sigma_{\rm [Fe/H]} \sim 0.4$~dex.
We also identified seven EMP candidates, while the MR end of the MD is populated by faint main-sequence stars with relatively larger uncertainties ($\delta{\rm [Fe/H]} \sim 0.5$~dex).

Our results represent a significant improvement on the spectroscopic literature (see Fig.~\ref{fig:Hist-FeH}), in which metallicities were derived for only a few dozen RGB stars, mostly concentrated in the inner part of the galaxy \citep[e.g.][]{Kirby2013, Sitnova2021}. We also note that deep photometric follow-ups have shown Coma~Berenices having a regular surface brightness profile with no signs of tidal perturbations \citep{Munoz2010,Munoz2018b}, with our most external members found within the galaxy nominal tidal radius \citep[see][again]{Munoz2018b}.

\paragraph{\textbf{Draco.}}
Draco is one of the brightest systems in our sample ($M_V \sim -9$) and one of the best characterised.
It hosts a predominantly old ($>10$~Gyr) and metal-poor stellar population \citep[e.g.][]{Aparicio2001,Kirby2011,Ding2025} with no clear signs of tidal disruption \citep[][]{Segall2007,Munoz2018b,Jensen2024}, although member stars have been found beyond its nominal tidal radius \citep[see][]{Ding2025}. 

We found 1\,339 probable members extending down to two magnitudes below the horizontal branch, as shown in Fig.~\ref{fig:FEH-dist-UFD-1} (bottom panel). Due to the extended apparent size of Draco, we only have a complete coverage within $3\times R_h$, with members found as far as $5\times R_h$. We observe a tentatively radial metallicity gradient in which metal-rich stars are more centrally concentrated than metal-poor stars, in line with recent spectroscopic results \citep{Ding2025}.

Compared to the spectroscopic samples in the literature (see Table~\ref{tab:sample}), we provide a relatively modest, yet crucial, increase in sample size, particularly in the metal-poor tail of the MD. We identified 62 EMP candidates, mostly with no spectroscopic measurements yet.
We also note the presence of two MR stars with ${\rm [Fe/H]} > -1.25$~dex, which are in very good agreement with their spectroscopic values \citep{Geha2026}. However, given their distance from the centre ($1<R/R_h<3$), we cannot rule out the possibility that they are actually MW contaminants.

\paragraph{\textbf{Hercules.}}
This UFD hosts an exclusively old stellar population ($>12$~Gyr), whose formation ended around the reionisation epoch \citep{Brown2014, Weisz2014}. Hercules's highly elongated morphology shows signs of tidal tails \citep[e.g.][]{Sand2009, Martin+Jin2010, Roderick2015}, although whether its dynamical state is in equilibrium remains a matter of debate \citep[e.g.][]{Gregory2020,Longeard2023, Ou2024}.

We uncovered 74 probable member stars distributed along the RGB. As shown in Fig.~\ref{fig:FEH-dist-UFD-2} (top panel), they nicely follow the system's elongated structure, with probable members found as far as $7\times R_h$. We also identified 8 EMP candidate stars.

Compared to the spectroscopic literature (Table~\ref{tab:sample}), new members are found at all radii, especially at the farthest ones. This demonstrates the effectiveness of our technique in identifying potential members that are deeply embedded within the contamination field. Their spectroscopic follow-up could prove crucial for verifying the tidal disruption scenario and clarifying Hercules's dynamical state.

\paragraph{\textbf{Leo~IV and Leo~V.}}
We discuss these two UFDs together, as they form a close pair in the sky ($d \sim 3$ degrees), but also because they share similar heliocentric distances ($\sim150$ kpc; Table~\ref{tab:sample}) and orbital properties \citep[e.g.][]{Battaglia2022}. These findings indeed suggest that they were accreted together as part of the Leo–Crater group \citep{Julio2024}.

Thanks to their limited extension, we covered both systems well beyond their nominal tidal radii \citep[$\sim10$~arcmin][]{Munoz2018b}, as shown in Fig.~\ref{fig:FEH-dist-UFD-2} (lower panels). We sampled their RGB down to the horizontal branch, finding 25 and 14 probable members respectively in Leo~IV and Leo~V. In particular for Leo~IV, all members are found within $3\times R_h$, while for Leo~V they extend up to $8\times R_h$. The latter might provide evidence of tidal disturbance in the outer regions of Leo~V \citep[e.g.][]{Sand2012, Collins2017, Mutlu-Pakdil2019}. We also find one EMP candidate in Leo~IV. 

Compared to the spectroscopic literature, the improvement is modest (see Fig.~\ref{fig:Hist-FeH}), meaning that we are likely observing most of the member stars down to $g_0 \sim 23$. While there is general agreement between the observed and spectroscopic MDs in Leo~V, in Leo~IV we see a clear offset between the modes of the two distributions.  
This is caused by a subsample of stars whose values deviate from one another by $\approx0.5$~dex.

It is hard to understand the origin of the offset in Leo~IV, given the general agreement with the literature that we observe for the other systems. Our measurements show a uniform distribution on the CaHK calibration plane, with no sign of problematic deviations. We find excellent agreement with the high-resolution studies of \citet{Simon2010} and \citet{Francois2016}, but much less so with those of lower resolution \citep[][]{Jenkins2021,Geha2026b}.
Finally, our MD shows the typical negatively skewed shape, unlike the spectroscopic one. Clearly, further observations of this system are needed to draw a definitive conclusion.

\paragraph{\textbf{Pisces~II.}}
One of the faintest and most distant system in our sample (see Table~\ref{tab:sample}), Pisces~II appears extremely compact in the sky. This allowed us to cover it well beyond its nominal tidal radius \citep[$\sim8$~arcmin][]{Munoz2018b}, as shown in Fig.~\ref{fig:FEH-dist-UFD-3} (top panel).

We found 11 probable member stars along the RGB and extending down to the horizontal branch. These stars have an extended spatial distribution, reaching as far as $6\times R_h$. 
The MD shows a clear peak at ${\rm [Fe/H]} \simeq -2.4$~dex, while the large standard deviation ($\sigma_{\rm [Fe/H]} \simeq 0.7$~dex) is due to the fainter and more uncertain stars (with $\delta{\rm [Fe/H]} > 0.2$~dex) at the extremes of the distribution. The two EMPs we have identified are found among these.

The comparison with the spectroscopic literature is good, with a small 0.1~dex offset between common values (see Fig.~\ref{fig:Hist-FeH}).
Taking into account stars with radial velocity measurements as well, we add only one new probable member without previous spectroscopic data, which means that our selection is likely to be complete up to $g_0\sim23$. 
We also note that a spectroscopically confirmed CEMP-no star in Pisces~II \citep{Spite2018} is excluded from our analysis, as its red colour $(g-r)_0\sim1.0$ places it outside our photometric selection window.

If we were to exclude the MD extreme values, the result would be a very narrow distribution ($\sigma_{\rm [Fe/H]} \sim 0.2$~dex), which would lead us to question the galactic nature of Pisces~II. To verify that it is not a globular cluster, we calculate its velocity dispersion with the remaining stars \citep[following][]{Taibi2018}. We obtain a value of $\sigma_{\rm v} = 3.8^{+2.1}_{-1.3}$~km\,s$^{-1}$, consistent with values reported in the literature \citep[][]{Kirby2015,Geha2026}. However, given the large uncertainties involved, a deeper spectroscopic follow-up of Pisces~II is necessary to clearly establish that it is an UFD.

\paragraph{\textbf{Segue~1.}}
This UFD is the faintest system in our sample, but also the closest one (see Table~\ref{tab:sample}). We were able to deeply cover its extent well beyond its nominal tidal radius \citep[$\sim8$~arcmin][]{Munoz2018b}, as shown in Fig.~\ref{fig:FEH-dist-UFD-3} (central panel).

We found 29 probable members down to more than two magnitudes along the main sequence, significantly improving the available spectroscopic metallicities (Fig.~\ref{fig:Hist-FeH}). Probable members are found out to $\sim 5 \times R_h$, with the two most radially distant stars being spectroscopically confirmed members \citep{Simon2011, Frebel2014}. 

The MD shows large dispersion ($\sigma_{\rm [Fe/H]} \simeq 0.7$~dex) with a possible bimodal structure, peaking at ${\rm [Fe/H]} \simeq -2.0$~dex and at $\simeq -3.0$~dex. Although this is not the first case observed among the UFDs of a bimodal distribution \citep[i.e. Reticulum~II,][]{Luna2025}, the MD of Segue~1 remains difficult to reconcile with its short star formation history \citep{Durbin2025}, unless a recent burst of star formation is invoked \citep[see again][]{Luna2025}. 

A more plausible explanation is the presence of contaminants. Indeed, Segue~1 overlaps along the line of sight with the Sagittarius stream and 300S, a fainter stream discovered as a kinematic overdensity at around 300~km\,s$^{-1}$ \citep[e.g.][]{Niederste-Ostholt2009,Simon2011,Fu2018}. These streams could contribute to the observed distribution with metal-rich contaminants (i.e. with $\rm [Fe/H]>-2.0$~dex).

All but one of our probable members on the MR peak have measured radial velocities, which are within $\pm10$~km\,s$^{-1}$ of the systemic radial velocity of Segue~1 of $\sim200$~km\,s$^{-1}$ \citep {Simon2011,Geha2026}. This clearly separates them from the 300S stream. 

On the other hand, their association with the Sagittarius stream seems more plausible. From the \citet{Vasiliev2021} model of the stream, we find that Segue~1 shares with it both location on the sky, and similar 3D velocity and distance. Moreover, the Sagittarius stream stars selected from \citet{Ramos2022} lie in the same region of the CMD as the Segue~1 RGB and exhibit comparable metallicities (i.e. ${\rm [Fe/H]}\sim-1.5$~dex).

Although it is hard to establish which RGB star in Segue~1 is a contaminant, we can assume that if they all belonged to Sagittarius, the other MR stars found on the main sequence could contribute to a spurious metal-rich peak due to their larger metallicity uncertainties ($>0.2$ dex). Clearly, a deep spectroscopic follow-up study is needed to resolve this issue.

\paragraph{\textbf{Ursa~Major~I.}}
This relatively distant system ($D_\odot\sim100$~kpc, see Table~\ref{tab:sample}) appears to be a fossil galaxy, hosting an old, metal-poor population, as do the other UFDs in our sample \citep[e.g.][]{Brown2014}. However, its morphology is highly elongated and disturbed, showing signs of tidal stripping \citep{Martin2008,Okamoto2008,Munoz2018b}.

We identified 51 probable member stars along the RGB down to almost the main sequence turn-off. Their spatial distribution follow the system's morphology, with probable members found as far as $4\times R_h$. 
The MD shows a clear peak at ${\rm [Fe/H]} \simeq -2.5$~dex, with a long metal-poor tail containing 10 EMP candidates.

The observed MD closely resembles the spectroscopic one, although we double the number of observed stars.
Most of the new members that we have identified are metal-poor stars located beyond $2\times R_h$, revealing a new population that, if spectroscopically confirmed, could provide new insights into the dynamical status of Ursa~Major~I.

We note that two MR stars clearly detach from the rest of the MD population (see Fig.~\ref{fig:FEH-dist-UFD-3}). Their location on the CaHK calibration plane is close to the contaminants locus, while spatially they are close to the galaxy's centre. A comparison with available spectroscopic data \citep{Kirby2013, Geha2026} confirms that one of them is indeed metal-rich, while the other has a $\sim 0.5$~dex lower value closer to the system's average metallicity. 

\paragraph{\textbf{Ursa~Major~II.}}
Ursa~Major~II is a nearby UFD ($D_\odot \sim 35$~kpc, see Table~\ref{tab:sample}) that we continuously covered only within $3\times R_h$, as shown Fig.~\ref{fig:FEH-dist-UFD-4} (top panel). Its proximity enabled us to identify 185 potential members along the RGB extending down to two magnitudes below the main sequence turn-off. This is the highest number among the UFDs in our sample (see Table~\ref{tab:FEH-dist}).

The MD of probable member stars shows a peak at ${\rm [Fe/H]} \simeq -2.3$~dex. Compared to the spectroscopic literature, we have increased the number of metallicity estimations by nearly ten times (see Fig.~\ref{fig:Hist-FeH}). In fact, previous studies only probed the RGB, whereas we have extended the analysis to the main sequence. Note that most of these new additions have an associated radial velocity, reinforcing their membership status.

We identified four EMP candidates. This represents a relatively small fraction ($\sim 2\%$) compared to other UFDs in our sample ($\sim 5\%$). Conversely, the MR tail of the distribution is more populated, consisting largely of main sequence stars. As reported in Sect.~\ref{subsec:photo-met-cat}, the relatively large photometric uncertainties and reduced metallicity separation in this region of the calibration plane could introduce a bias towards higher metallicity estimates. 

One MR star is clearly detached from the distribution of the other members. Its location on the calibration plane, coupled with its relatively large distance from the centre ($\sim2.5\times R_h$), makes it a potential MW contaminant.
We also note that Ursa~Major~II may be subject to significant tidal disruption \citep{Munoz2010, Smith2013}. However, the limited spatial coverage restricts our ability to detect extra-tidal members.

\paragraph{\textbf{Ursa~Minor.}}
Ursa~Minor is one of the three brightest systems in our sample (see Table~\ref{tab:sample}). Similar to Draco, the covered area was limited to a radius of $3\times R_h$ due to its large extent across the sky, as shown in Fig.~\ref{fig:FEH-dist-UFD-4} (central panel). 
However, we note that Ursa~Minor is highly elongated and irregular at large radii, with extra-tidal features suggesting a significant tidal processing and, perhaps, a past major merger \citep[e.g.][]{Bellazzini2002, Sestito2023, Sato2025a, Sato2025b}.

We could identify 1\,710 probable member stars down to two magnitudes below the horizontal branch, resulting in the highest number among our galaxy sample. Their radial distribution shows a tentative metallicity gradient, given the limited spatial coverage. Nevertheless, the approximate $-0.1$~dex\,$R_h^{-1}$ decrease we measure is in line with previous spectroscopic measurements \citep{Pace2020,Taibi2022}.

Compared to the spectroscopic literature (see Table~\ref{tab:sample}), we were able to more than double the number of metallicity estimations (see also Fig.~\ref{fig:Hist-FeH}). The improvement is particularly relevant on the metal-poor tail of the MD, where the majority of the 51 EMPs we uncovered have had no spectroscopic follow-up so far.

\paragraph{\textbf{Willman~1.}}
Willman~1 is a nearby system ($D_\odot \sim 40$~kpc), among the faintest and most compact of our sample (see Table~\ref{tab:sample}). Although there is some evidence for its classification as a galaxy \citep{Martin2007, Willman2011}, its elongated structure and orbit make tidal disturbances plausible, possibly inflating its velocity dispersion \citep{Chiu2026arXiv}.

We found 29 probable members reaching one magnitude below the main sequence turn-off. Their spatial distribution follows the system's morphology, with members found as far as $4\times R_h$, as shown in Fig.~\ref{fig:FEH-dist-UFD-4} (bottom panel). Their MD shows a peak at ${\rm [Fe/H]} \simeq -2.3$~dex and a large spread $\sigma_{\rm [Fe/H]} \simeq 0.5$~dex.

Compared with the spectroscopic literature, we have more than tripled the number of metallicity values (see Fig.~\ref{fig:Hist-FeH}), mainly thanks to the probable members on the main sequence. 
However, the sample increase is concentrated mainly in the metal-rich part of the MD. This is once again attributable to the main sequence stars, consistent with the behaviour observed in Coma~Berenices and Ursa~Major~II.

\subsection{Luminosity-metallicity relation}
\label{subsec:lum-feh}

\begin{figure*}
    \centering
    \includegraphics[width=\textwidth]{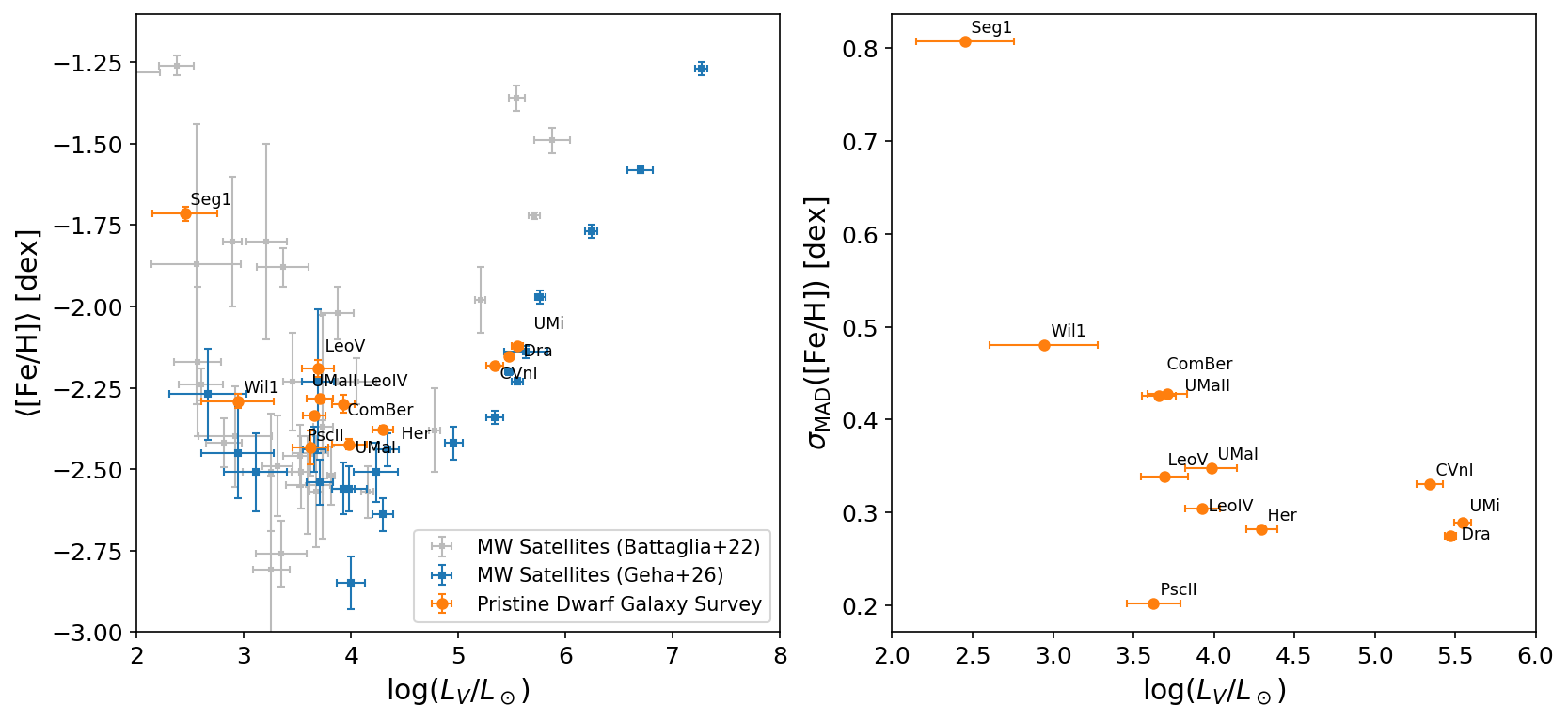}
    \caption{
    Luminosity-metallicity relation. In the \textit{left} panel, the error-weighted mean metallicity is shown as a function of luminosity for our sample (orange circles), the MW satellites from the \citet{Geha2026b} compilation (blue squares), and other MW systems from the \citet{Battaglia2022} compilation (grey circles). 
    In the \textit{right} panel, the measured metallicity dispersion (as median absolute deviation) for our sample. 
    }
    \label{fig:Lv-FeH}
\end{figure*}

Dwarf galaxies follow a well-known luminosity-metallicity relation, which is linear down to $L_V\sim10^5L_\odot$ \citep[e.g.][]{Dekel+Silk1986,Simon+Geha2007,Kirby2013,Simon2019}. For fainter systems (i.e. those in the UFD regime), average [Fe/H] measurements show increased scatter, which has also been interpreted as a metallicity floor at ${\rm [Fe/H]}\approx -2.5$~dex \citep{Simon2019,Fu2023,Geha2026b}. 

In the left panel of Fig.~\ref{fig:Lv-FeH}, we show the $L_V-{\rm [Fe/H]}$ relation for our sample of dwarf galaxies, compared to other MW satellites. Literature values are taken mainly from the \citet{Geha2026b} compilation, and from \citet{Battaglia2022} for the remaining systems, although in the latter case they have been compiled from a heterogeneous set of spectroscopic studies.

For the bright systems in our sample ($L_V>10^5L_\odot$), we find overall agreement with the linear trend of luminous dwarf galaxies \citep[see e.g.][]{Kirby2013, Geha2026b}. In contrast, the faint systems ($L_V<10^5L_\odot$) deviate from this trend, scattering around ${\rm [Fe/H]\sim-2.3}$~dex. This matches previous results \citep[e.g.][]{Fu2023,Geha2026b}, though our systems generally have $\sim0.1$~dex higher mean metallicities, excluding Segue~1 for the reasons noted above (see also Appendix~\ref{apx:results} for a comparison with the \citet{Geha2026} compilation). Therefore, our analysis confirms a deviation from the linear $L_V-{\rm [Fe/H]}$ relation in the UFD regime, after accounting for a possible general offset.

In the right panel of Fig.~\ref{fig:Lv-FeH}, we show the measured metallicity dispersion as median absolute deviation $\sigma_{\rm MAD}(\rm [Fe/H])$ (see Table~\ref{tab:FEH-dist}) of our systems as a function of their luminosity. We observe an increase of the scatter of values towards fainter luminosities, with Segue~1 being again a significant outlier. The observed trend is consistent with previous findings \citep[e.g.][]{Fu2023,Geha2026b}. If physical, it might be attributed to the stochastic nature of chemical enrichment processes and metal mixing during galaxy evolution. An important difference in our case is that we do not assume a specific functional form for the metallicity distribution (such as a Gaussian, as is usually the case), but instead use a robust scale indicator that should yield better results for an underlying heavy-tailed distribution \citep[see][]{Beers1990}.

The presence of a metallicity floor at the UFD level, on the other hand, is difficult to reconcile with model predictions \citep[e.g.][]{Jeon2017,Jeon2021,Applebaum2021,Sanati2023}. Although EMP candidates are indeed found in UFD systems, the overall level of chemical enrichment appears to be similar across these systems, even when their masses are very low and, in principle, the total mass of gas available for star formation -- and thus for subsequent chemical enrichment -- should decrease accordingly.

Recent hydrodynamic simulations and semi-analytical models of MW-like halos \citep[e.g.][]{Ahvazi2024,Rey2025arXiv} have reproduced the observed scatter at ${\rm [Fe/H]}\approx-2.5$~dex. However, it remains unclear whether this trend depends on external factors, such as the composition of the intergalactic medium during UFD formation \citep{Ahvazi2024}, or from internal enrichment by the first stars \citep{Sanati2023,Rey2025arXiv}. More UFD observations are needed to confirm the metallicity floor and constrain the models.

\subsection{Radial metallicity gradients}

\begin{figure}
    \centering
    \includegraphics[width=.49\textwidth]{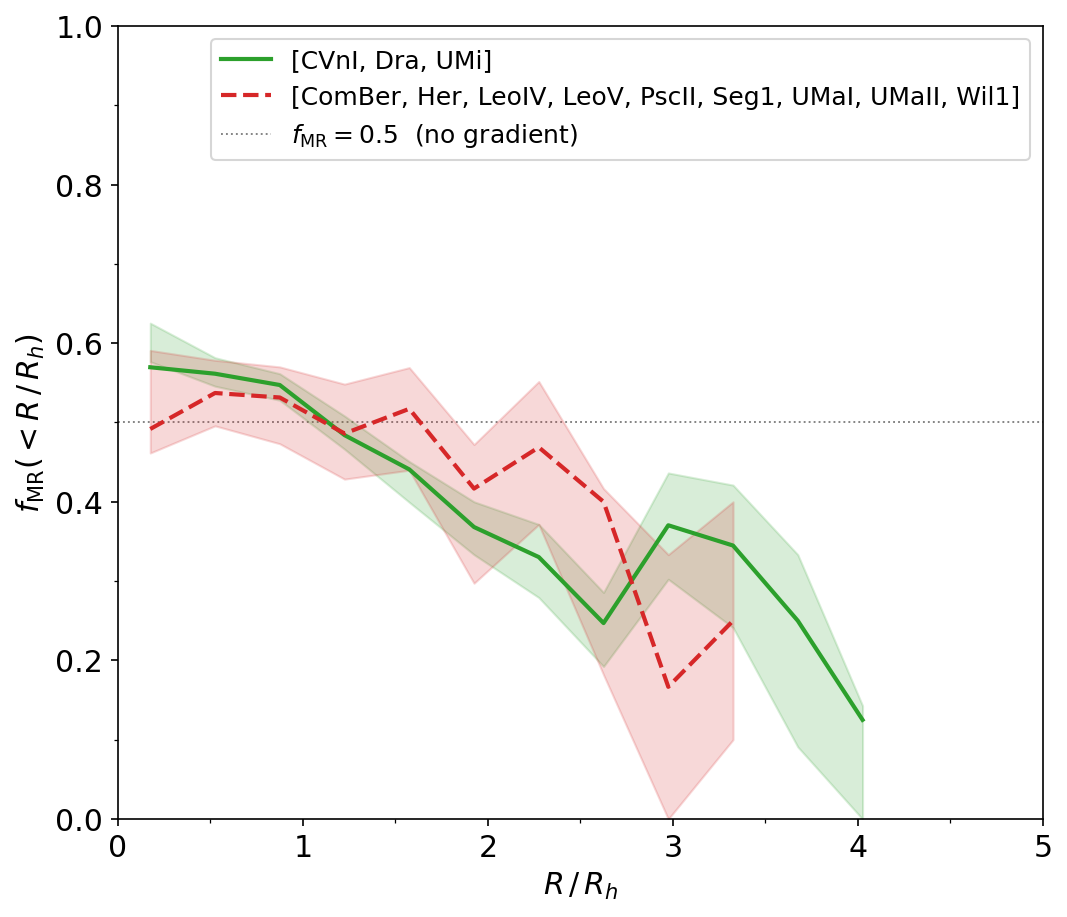}
    \caption{Cumulative fraction of metal-rich stars as a function of the radial distance scaled to the half-light radius of each system. Metal-rich stars here are those with a [Fe/H] value greater than the median of the respective system. The green (red) solid (dashed) line represents the dSph (UFD) systems in our sample. The shaded bands represent the 1$\sigma$ confidence intervals.}
    \label{fig:Rad-FeH}
\end{figure}

Radial metallicity gradients are common among dwarf galaxies \citep[e.g.][]{Taibi2022}. They can form secularly due to internal star formation processes \citep[e.g.][]{Pontzen2014,Revaz+Jablonka2018}, with external factors such as specific merger events having the ability to strengthen the gradients \citep[e.g.][]{Benitez-Llambay2016}.
However, UFDs are not expected to exhibit a prominent metallicity gradient, due to their short star formation history (SFH) and stochastic chemical enrichment, combined with the dynamical mixing of different stellar populations \citep[e.g.][]{Revaz+Jablonka2018}. 

In our sample, a radially declining [Fe/H] profile is clearly visible in Canes~Venatici~I, Draco, and Ursa~Minor, although the latter two were surveyed only within $3 \times R_h$ (see Figs.~\ref{fig:FEH-dist-UFD-1}-\ref{fig:FEH-dist-UFD-4}), so their outer profiles remain unconstrained. On the other hand, such trends are more tentative among the remaining systems.
To make the analysis more robust, we aggregate the radial [Fe/H] distributions of either the bright or the remaining faint systems. These are split according to the median [Fe/H] values of each system into a metal-richer and a metal-poorer sample. Performing a two-sample Kolmogorov–Smirnov (KS) test, we recover a statistically significant difference only for the bright systems ($D=0.15$ and $p=3.3\times10^{-17}$), while the faint systems show no significant separation ($p>0.05$).

This is further illustrated in Fig.~\ref{fig:Rad-FeH}, which shows the aggregate fraction of metal-rich stars, $f_{\rm MR}$, in the dSphs and UFDs as a function of radial distance, scaled to each system's half-light radius $R_h$. Metal-rich stars in this case are defined as those with a [Fe/H] value greater than the median of the respective system. The shaded bands represent the $1\sigma$ (68\%) confidence intervals estimated by bootstrapping for each sample, performing 1000 resamples with replacement from the stellar catalogue and recomputing $f_{\rm MR}$ at each radial bin. 

In Fig.~\ref{fig:Rad-FeH} we observe indeed a significant decline for the bright systems, while the faint systems show no significant trend within $2.5 \times R_h$, with a tentative decline only afterwards. However, its significance is low considering the small number of stars in the outskirts. We also cannot exclude a selection effect arising from membership probability estimation; deeper data would be needed to draw a conclusive statement on the faint systems.

%%%%%%%%%%%%%%%%%%%%%%%%%%%%%%%%%%%%%%%%%%%%%%%%%%%%%%%%%%%%%%
\section{Conclusions}
\label{sec:conclusions}

In this work, we presented a homogeneous photometric metallicity catalogue for a sample of 12 Milky Way faint satellites, spanning a luminosity range from the classical dwarf spheroidals down to the ultra-faint dwarf regime ($-9.0 \lesssim M_V \lesssim -1.3$). 
Our analysis combined deep \textit{CaHK} narrow-band photometry from the \textit{Pristine} dwarf galaxy survey with broad-band $g$ and $r$ data from \citet{Munoz2018} and Pan-STARRS1, enabling us to reach unprecedented photometric depth and spatial extent across all systems. 

We presented an improved photometric metallicity method based on a new calibration grid defined in the $(CaHK-g)_0 - 2.4\times(g-r)_0$ vs $(g-r)_0$ colour-colour plane, using $\sim19\,500$ calibration stars covering both the giant and main sequence regimes. This has enabled us to assign photometric metallicities to stars ranging from the red giant branch to the main sequence in our target systems.

Using an updated version of the probabilistic membership method of \citet{Battaglia2022}, we identified a total of 3\,917 probable member stars ($\mathcal{P}_\mathrm{f} \geq 0.8$) across the 12 systems, more than doubling the number of members recovered in previous comparable analyses. Probable members were found out to radial distances of $5$--$8 \times R_h$ in several systems, demonstrating the power of photometric surveys for mapping stellar populations in the poorly explored outskirts of dwarf galaxies.

We performed extensive internal and external validation of the photometric metallicities. Comparisons with high-resolution spectroscopic compilations, along with the \citet{Walker2023}, DESI-DR1 \citep{Koposov2026}, and \citet{Geha2026} catalogues, consistently show no significant biases, with a scatter of $\sim0.20$~dex on the residuals. 
We have, however, detected small systematic deviations of $\sim0.1$~dex on a system-by-system basis when compared with the \citet{Geha2026} compilation.

We obtained reliable metallicity distributions, providing robust mean metallicities and dispersions that were previously inaccessible for most systems in this study. The three brightest systems (Canes~Venatici~I, Draco, and Ursa~Minor) show well-defined distributions with a clear peak, moderate dispersions ($\sigma_\mathrm{[Fe/H]} \sim 0.35$--$0.40$~dex), and a negatively skewed tail to lower metallicities. Among the UFDs, most have similar dispersions despite their lower luminosities, except for Segue~1 and Pisces~II, whose broader distributions may result from foreground contamination or limited member samples.

We identified 170 candidate extremely metal-poor (EMP) stars with $\mathrm{[Fe/H]} < -3.0$~dex, found in all of the surveyed systems. These EMP candidates cover the full magnitude range spanned by our sample of probable member stars. Although the selection purity is expected to decline at fainter magnitudes, this catalogue represents a valuable list of targets that must be robustly confirmed through spectroscopic follow-up.

The brighter systems in our sample follow the well-established luminosity-metallicity relation. The UFDs, however, depart from the extrapolated trend, scattering around $\mathrm{[Fe/H]} \sim -2.3$~dex with no clear dependence on luminosity. This is consistent with a metallicity floor in the UFD regime, as reported in previous studies, and remains a challenge for theoretical models of galaxy formation and chemical enrichment.

Statistically significant radial metallicity gradients were detected in Canes~Venatici~I, Draco, and Ursa~Minor, with metal-rich stars being more centrally concentrated than metal-poor ones. No significant gradient was found among the UFDs within $2.5 \times R_h$, in line with expectations from their short star formation histories and stochastic enrichment. A tentative decline at larger radii was observed, but remains inconclusive given the limited number of stars in the outermost regions.

Our results show that photometric metallicity techniques constitute an effective and complementary approach to spectroscopic studies, overcoming the latter’s limitations in terms of depth and spatial coverage. While wide-field spectroscopic surveys are currently underway (e.g. WEAVE, 4MOST, MOONS, PFS), these will explore the ultra-faint regime only to a limited extent, and deep photometric surveys (e.g. UNIONS, LSST, Euclid) combined with the use of narrow-band filters will continue to be an essential complement for several years to come.

\section*{Acknowledgements}

This work was supported by the Swiss National Science Foundation (SNSF) under funding reference 200021-213076 “Galaxy evolution in the cosmic web”. 
This research was supported by the International Space Science Institute (ISSI) in Bern, through ISSI International Team project 540 (The Early Milky Way). The authors thank the International Space Science Institute, Bern, Switzerland, for providing financial support and meeting facilities to the international team Pristine. 
This work was supported by the Programme National Astro of CNRS/INSU with INP and IN2P3, co-funded by CEA and CNES.
GB acknowledges support from the Agencia Estatal de Investigación del Ministerio de Ciencia, Innovación y Universidades (MCIU/AEI) under grant “EN LA FRONTERA DE LA ARQUEOLOGÍA GALÁCTICA: EVOLUCIÓN DE LA MATERIA LUMINOSA Y OSCURA DE LA VÍA LÁCTEA Y LAS GALAXIAS ENANAS DEL GRUPO LOCAL EN LA ERA DE GAIA (FOGALERA)” and the European Regional Development Fund (ERDF) with reference PID2023-150319NB-C21 / 10.13039/501100011033.
AA acknowledges support from the Herchel Smith Fellowship at the University of Cambridge.
FS acknowledges funding from the UK Science and Technology Facilities Council through grant ST/Y001443/1.
Based on observations obtained with MegaPrime/MegaCam, a joint project of CFHT and CEA/DAPNIA, at the Canada-France-Hawaii Telescope (CFHT) which is operated by the National Research Council (NRC) of Canada, the Institut National des Sciences de l’Univers of the Centre National de la Recherche Scientifique of France, and the University of Hawaii. 
The Pan-STARRS1 Surveys (PS1) and the PS1 public science archive have been made possible through contributions by the Institute for Astronomy, the University of Hawaii, the Pan-STARRS Project Office, the Max-Planck Society and its participating institutes, the Max Planck Institute for Astronomy, Heidelberg and the Max Planck Institute for Extraterrestrial Physics, Garching, The Johns Hopkins University, Durham University, the University of Edinburgh, the Queen's University Belfast, the Harvard-Smithsonian Center for Astrophysics, the Las Cumbres Observatory Global Telescope Network Incorporated, the National Central University of Taiwan, the Space Telescope Science Institute, the National Aeronautics and Space Administration under Grant No. NNX08AR22G issued through the Planetary Science Division of the NASA Science Mission Directorate, the National Science Foundation Grant No. AST–1238877, the University of Maryland, Eotvos Lorand University (ELTE), the Los Alamos National Laboratory, and the Gordon and Betty Moore Foundation.
This work has made use of data from the European Space Agency (ESA) mission \textit{Gaia} (\url{https://www. cosmos.esa.int/gaia}), processed by the \textit{Gaia} Data Processing and Analysis Consortium (DPAC, \url{https://www.cosmos.esa.int/web/gaia/dpac/ consortium}).
This research has made use of NASA’s Astrophysics Data System, VizieR catalogue access tool (CDS, Strasbourg, France, DOI: 10.26093/cds/vizier), and extensive use of python3.13 \citep{Python3}, including ipython \citep[v9.7,][]{ipython}, numpy \citep[v2.3,][]{NumPy-Array}, scipy \citep[v1.16,][]{SciPy-NMeth}, matplotlib \citep[v3.10,][]{Matplotlib}, pandas \citep[v2.3,][]{Pandas}, and astropy \citep[v7.1,][]{Astropy}.

%%%%%%%%%%%%%%%%%%%%%%%%%%%%%%%%%%%%%%%%%%%%%%%%%%
\section*{Data Availability}

The data underlying this article will be shared on reasonable request to the corresponding author.



\bibliographystyle{mnras}
\bibliography{biblio} % if your bibtex file is called example.bib

%\begin{thebibliography}{99}
%\end{thebibliography}

%%%%%%%%%%%%%%%%%%%%%%%%%%%%%%%%%%%%%%%%%%%%%%%%%%

%%%%%%%%%%%%%%%%% APPENDICES %%%%%%%%%%%%%%%%%%%%%

\appendix

\section{Samples -- Calibration}
\label{apx:sample}

\begin{figure*}
    \centering
    \includegraphics[width=.9\textwidth]{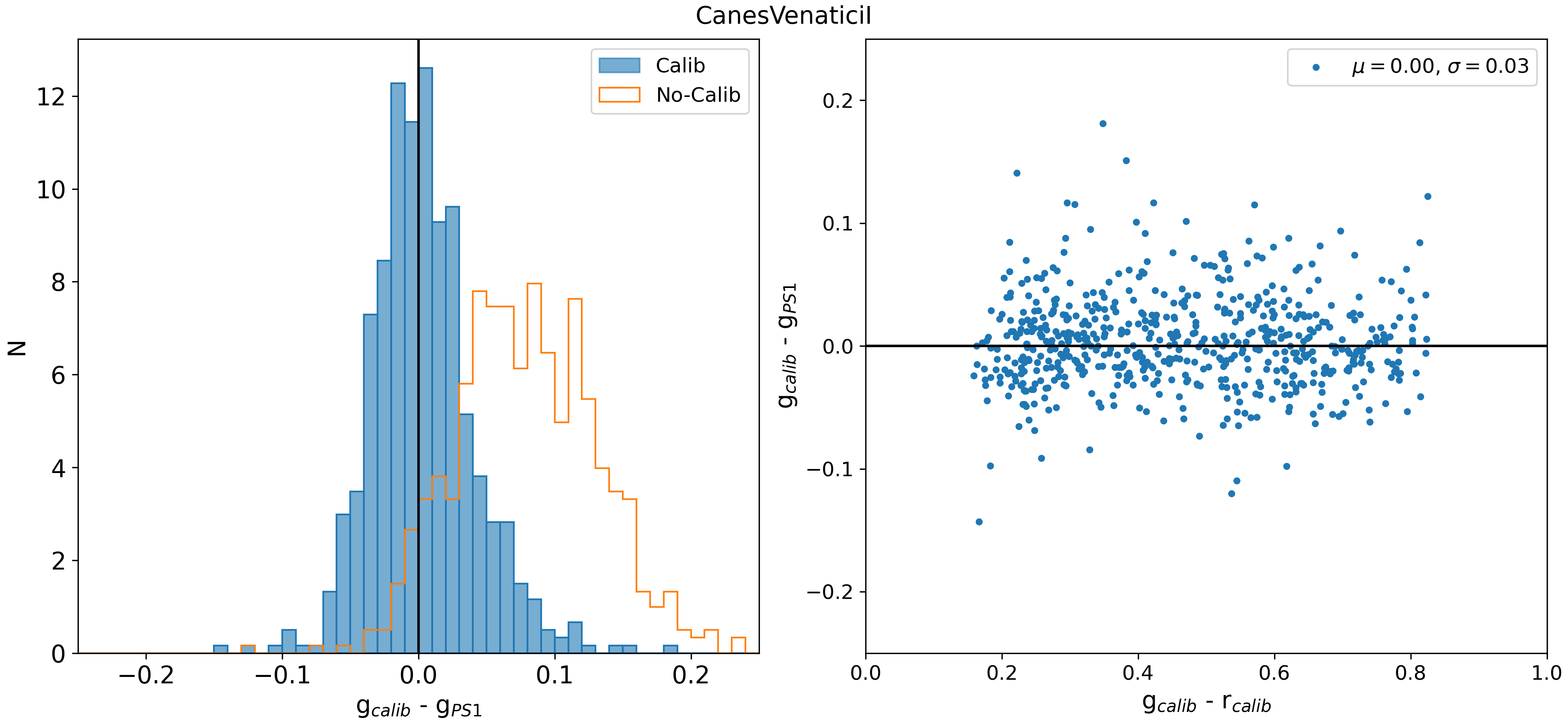}\\
    \includegraphics[width=.9\textwidth]{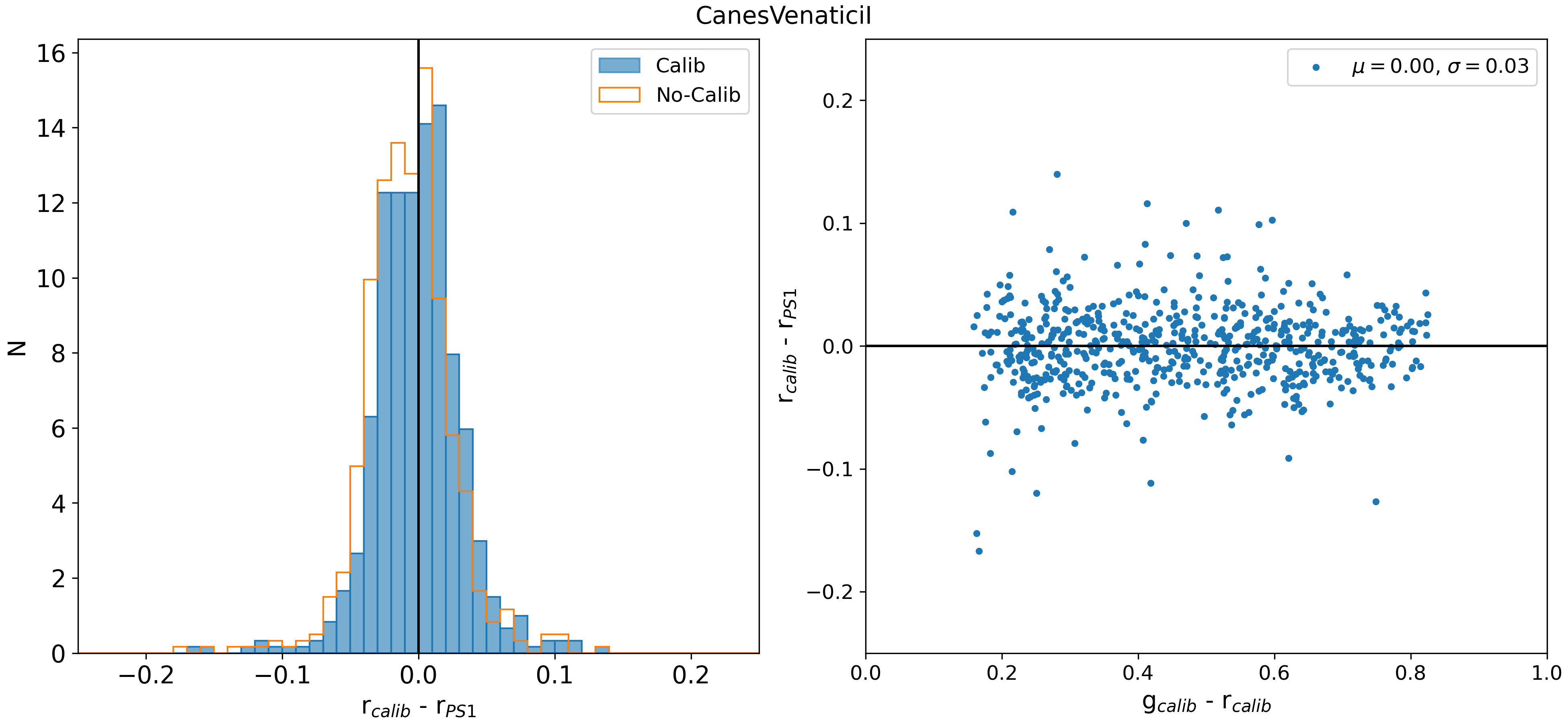}\\
    \includegraphics[width=.47\textwidth]{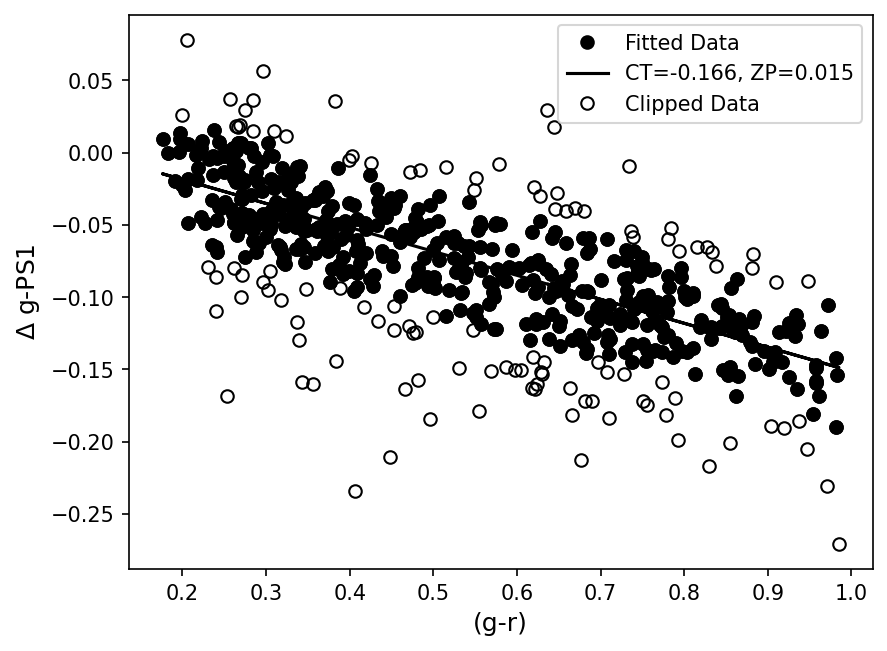}
    \includegraphics[width=.47\textwidth]{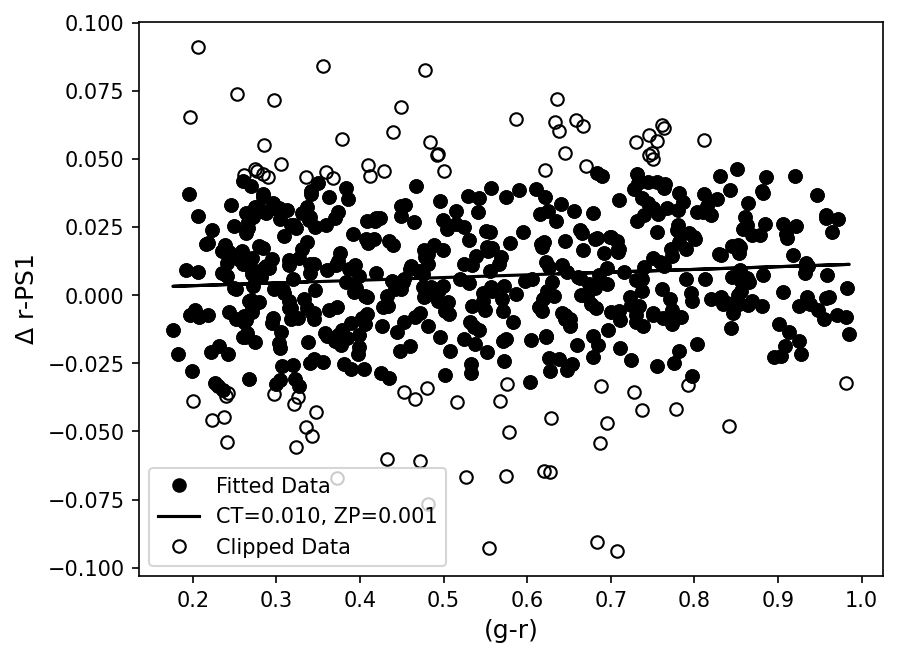}
    \caption{Calibration of the M18 photometry of Canes~Venatici~I based on the PS1 values of common stars. Top and middle panels show the histograms and color trend of the residuals between the calibrated and reference magnitudes respectively for the \textit{g} and \textit{r} bands (median and scatter reported in the legend boxes). The bottom panels show instead the linear calibration applied for the two photometric bands (colour term and zero-point reported in the legend boxes).}
    \label{fig:PS1calib}
\end{figure*}

To calibrate M18 over PS1, we selected common stars with $18<g_{\rm PS1}<21.5$ and $\delta_{m, \rm PS1}<0.05$ in each dwarf galaxy field. We then derived the colour equations for each band by performing the following linear fit with a 2$\sigma$ clipping:
\begin{equation}
    m_{\rm PS1} = m_{\rm M18} + ZP_m + c_m(g-r)_{M18}
\end{equation}
where $m$ is either $g$ or $r$, $ZP_m$ is the zero-point, and $c_m$ is the colour term. 

Magnitudes were previously corrected for extinction by using the \citet{S+F2011} recalibration of the \citet*{SFD1998} dust maps. In particular, we used the extinction coefficients provided in \citet{S+F2011} for the PS1 and SDSS passbands.

The zero points $ZP_m$ and colour terms $c_m$ were found to differ slightly from one field to another, with residuals between calibrated and reference PS1 magnitudes averaging around zero, with a typical standard deviation of $0.02-0.03$~mag. In Fig.~\ref{fig:PS1calib} are shown the results for Canes~Venatici~I. 

\section{Methods -- Individual likelihood terms}
\label{apx:likelihood}

Here, we describe the individual terms of the likelihood $\mathcal{L}$ used to obtain membership probabilities, as outlined in Sect.~\ref{subsec:membership}. 
We note that our method is an updated version of that presented in \citet{Battaglia2022}, with key differences outlined below.
We recall the total likelihood:
\begin{equation}
    \mathcal{L}_{\rm sat/MW} = \mathcal{L}_{\rm Sp}\mathcal{L}_{\rm CMD}\mathcal{L}_{\rm PM}\mathcal{L}_{\rm RV}
\end{equation}
where each term was treated separately for both the satellite and contamination samples.

\subsection{Spatial likelihood}
For the spatial $\mathcal{L}_{\rm Sp}$ term of the satellite, we assumed an exponential surface density profile for all systems. We constructed it by co-adding and normalising 1\,000 Monte-Carlo realisations propagating over the uncertainties in half-light radius, position angle and ellipticity 
\citep[see again][their Table~B.1 parameter values]{Battaglia2022}.
The contamination $\mathcal{L}_{\rm Sp}$ was instead assumed to be a uniform distribution, assuming negligible any possible MW sub-structures in the relatively small FoV of each system.

\subsection{Colour-magnitude likelihood}

\begin{figure*}
    \centering
    \includegraphics[width=.49\textwidth]{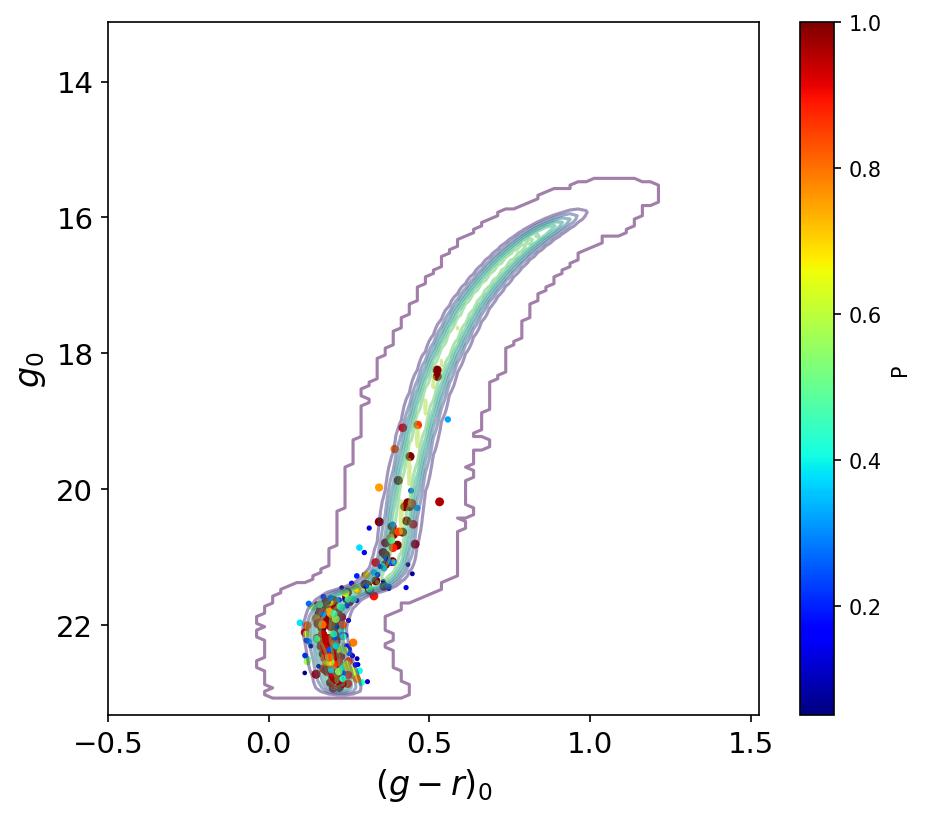}
    \includegraphics[width=.49\textwidth]{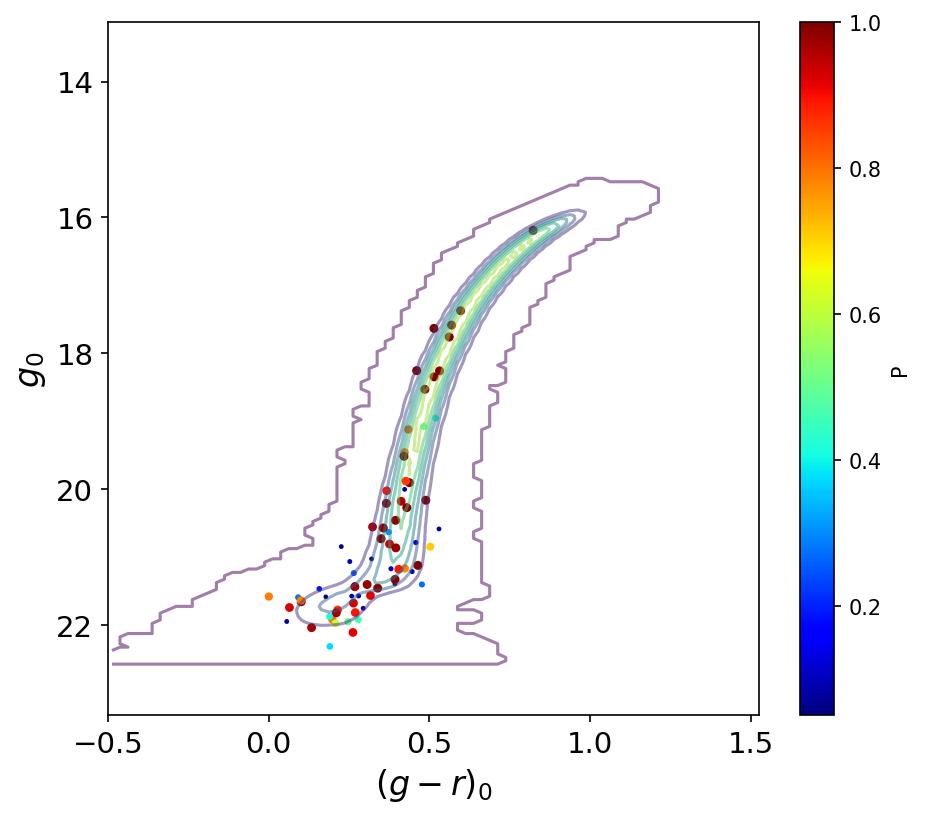}
    \caption{Colour-magnitude diagram of Coma Berenices using values from the M18 and PS1 catalogues respectively on the left and on the right, with the iso-contours of the photometric likelihood $\mathcal{L}_{\rm CMD}$ superimposed. The circles are coloured according to the calculated membership probabilities $\mathcal{P}_{\rm i}$.}
    \label{fig:Lcmd}
\end{figure*}

For the colour-magnitude $\mathcal{L}_{\rm CMD}$ term, we constructed the 2D look-up maps over a grid defined as follows: $-0.5<(g-r)<1.55$ and $13 < g < 23.5$~mag. We note that the $g$ limits cover the magnitude range of our data and the $(g-r)$ limits better isolate the loci of the RGB and MS in our systems. 

Unlike \citet{Battaglia2022}, we constructed the satellite $\mathcal{L}_{\rm CMD}$ term as in \citet{McConnachie+Venn2020}. For each system, we adopted a 13~Gyr Dartmouth isochrone \citep{Dotter2008} in the PS1 photometric bands, assuming an appropriate mean metallicity and distance modulus \citep[again from][their Table B.1]{Battaglia2022}, and excluding the post-He flash evolutionary phase. 

Similarly to the spatial term, we co-added and normalised 1\,000 realisations, moving the isochrone according to the distance modulus error. We accounted for the photometric uncertainties by assuming an exponential model obtained from the observed data as a function of magnitude. We also introduced an intrinsic colour spread of 0.03 mag. 

In order to take into account the photometric completeness and construct a realistic term, we multiplied the luminosity function at the faint end of the magnitude range by the fraction of observed stars in that range, as determined from the contamination sample. Results in form of iso-contours are shown in Fig.\ref{fig:Lcmd}, respectively for the M18 and PS1 case.

The contamination $\mathcal{L}_{\rm CMD}$ was obtained empirically from the observed data, selecting all those stars sufficiently further away from the dwarf galaxy. In practice, we included all stars within the FoV with an elliptical radius greater than $5\times R_h$, except for a few particularly extended systems, for which we reduced the radius to $4\times R_h$. Finally, this look-up map was normalised and smoothed using a Gaussian kernel of a size comparable to the photometric uncertainties.

\subsection{Proper motion likelihood}
For the proper motion likelihood $\mathcal{L}_{\rm PM}$, the satellite term was defined as a multivariate Gaussian \citep[e.g.][]{Pace+Li2019}. The systemic proper motion, along with its corresponding uncertainties and correlation terms, was adopted from \citet{Battaglia2022}. However, we did not account for an intrinsic dispersion term because it was smaller than the proper motion uncertainties.

As for the colour-magnitude term, the contamination $\mathcal{L}_{\rm PM}$ was obtained empirically using Gaia-DR3 data for each system. We selected all the stars with an elliptical radius from the centre of the target satellite between $5\times R_h$ and 2~deg. Also in this case, the look-up map was normalised and smoothed using a Gaussian kernel of a size comparable to the photometric uncertainties.

\subsection{Radial velocity likelihood}
Finally, the radial velocity $\mathcal{L}_{\rm RV}$ term of the satellite was assumed to be Gaussian, with the systemic velocity and velocity dispersion values taken from \citet{Battaglia2022}. 
For the contamination $\mathcal{L}_{\rm RV}$ term instead, we turned to the \textit{Gaia} Universe Model Snapshot \citep[GUMS,][]{Robin2012}. We downloaded GUMS catalogues within 2~deg of each system with the same cuts applied to the observed \textit{Gaia} catalogue, and modelled the normalised velocity histogram (ranging between $-500<v_{\rm los}[{\rm km\,s^{-1}}]<500$) as the sum of three Gaussian terms corresponding to the MW's halo, thin and thick disk components.

We note that not all stars with photometric information have a proper motion or a radial velocity measurement, particularly the faintest sources. We therefore ensured that these stars have no weight in the estimation of the  $\mathcal{L}_{\rm PM}$ and $\mathcal{L}_{\rm RV}$ terms by assigning them null values with arbitrarily large uncertainties.

\section{Results -- Validation of the photometric metallicities}
\label{apx:results}

In order to test the quality of the derived ${\rm [Fe/H]_{Pr}}$ values, we performed a series of internal and external comparisons with spectroscopic metallicity catalogues. 

\subsection{Internal comparisons}
\label{apx:int-comparison}

\begin{figure*}
    \centering
    \includegraphics[width=.98\textwidth]{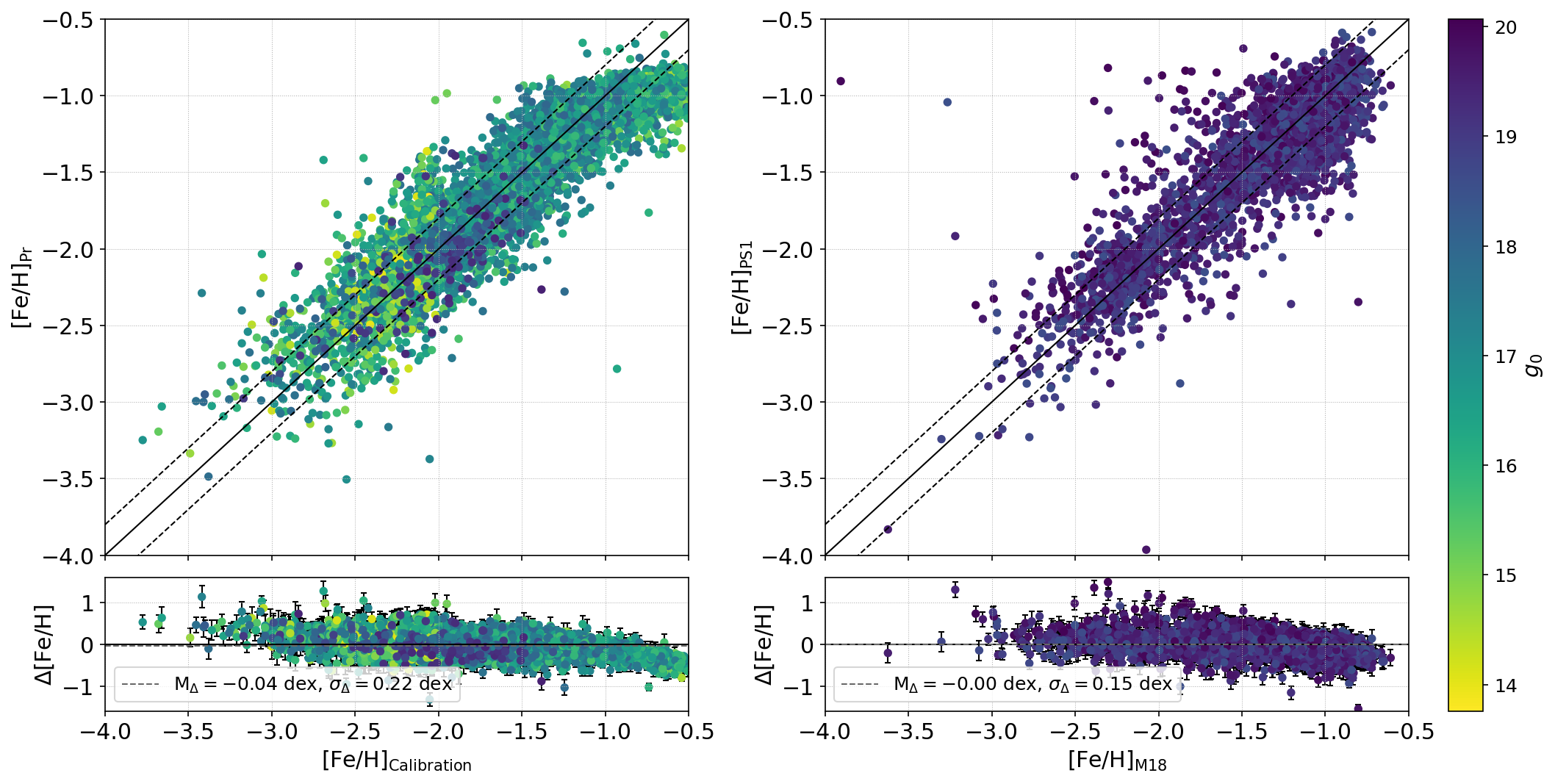} 
    \caption{
    Internal metallicity validation. The \textit{left} panel shows a comparison of the calculated metallicity values of individual stars in the calibration sample with spectroscopic values from the literature. The \textit{right} panel instead shows the comparison between the calculated photometric metallicities of common targets in the M18 and PS1 catalogues. The circles are coloured according to the stars $g_0$ magnitudes. The bottom panels also show the calculated median and median-absolute-deviation values of the residuals.}
    \label{fig:validation-int}
\end{figure*}

First, we calculated the photometric metallicities using the calibration sample to verify the method's internal consistency. Following \citet{Martin2024}, we further selected giant stars with ${\rm log}(g)<3.9$ and $T_{\rm eff}<6000$~K. In our case this also included dwarf galaxy values from the SAGA database and \citet{Walker2023} survey (see Sect.~\ref{subsec:photo-FeH}).
Additionally, we applied the metallicity quality cuts outlined in Sect.~\ref{subsec:photo-met-cat}, imposing though a limit of ${\rm [Fe/H]_{Pr,16th}}<-0.75$~dex and $\delta{\rm [Fe/H]_{Pr}}<0.2$~dex. 

As shown in the left panel of Fig.~\ref{fig:validation-int}, there is very good agreement between the calculated and spectroscopic metallicities for ${\rm [Fe/H]}<-1.0$~dex. In particular, the residuals generally show a small median offset ($<0.05$~dex) and scatter ($<0.25$~dex), with no significant dependency on $T_{\rm eff}$ or magnitude (spanning a range $14<g_0<19.5$ and $0<(g-r)_0<1$). This applies to the entire sample, as well as when the analysis is restricted to the SAGA and \citet{Walker2023} data only.

An exception is the subset of high-resolution spectroscopic follow-ups of Pristine \citep[see Sect. 7.4.1 in][and references therein]{Martin2024}, which shows a larger dispersion ($\sim0.5$~dex), mainly due to their relatively high effective temperatures ($T_{\rm eff}\sim6000$~K), where photometric metallicities are less reliable. Despite its small size ($\sim100$ stars), this subsample has been crucial for improving the low-metallicity statistics of the calibration set. Our comparison thus agrees with \citet{Martin2024}, although we use a different broadband photometric system (PS1 instead of Gaia).

Another important comparison concerned the main photometric metallicity catalogue presented in Sect.~\ref{subsec:photo-met-cat}. Since we use a combined M18 and PS1 catalogue, we verified that the common targets between the two sources provide consistent metallicities. We applied the standard metallicity quality cuts described above, while also restricting the magnitude range to $18.5<g_0<20.0$ for both catalogues. As shown in the right panel of Fig.~\ref{fig:validation-int}, values compare well with each other, with residuals having a median and dispersion comparable to the previous case.

\subsection{External comparisons}
\label{apx:ext-comparison}

\begin{figure*}
    \centering
    \includegraphics[width=.49\textwidth]{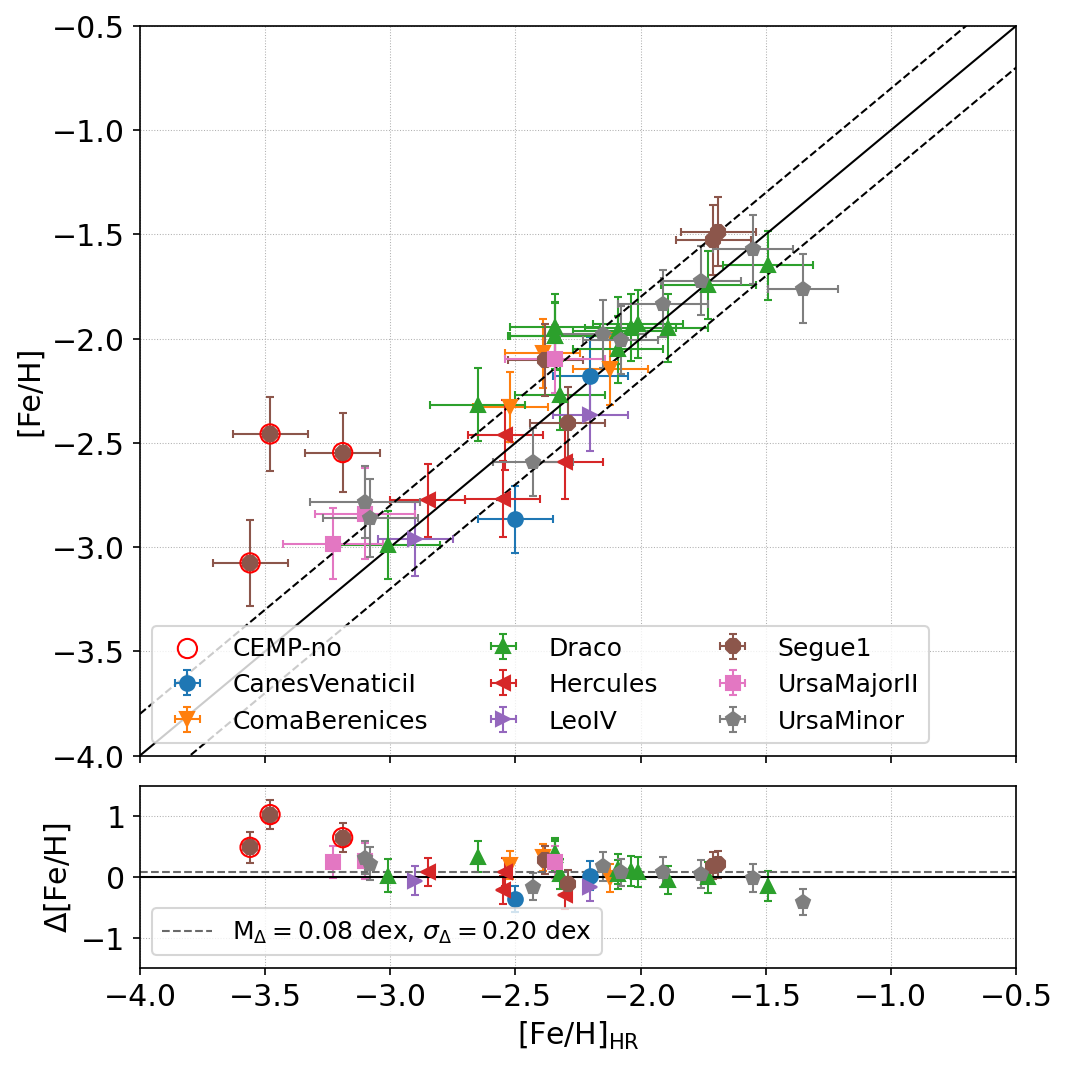}
    \includegraphics[width=.49\textwidth]{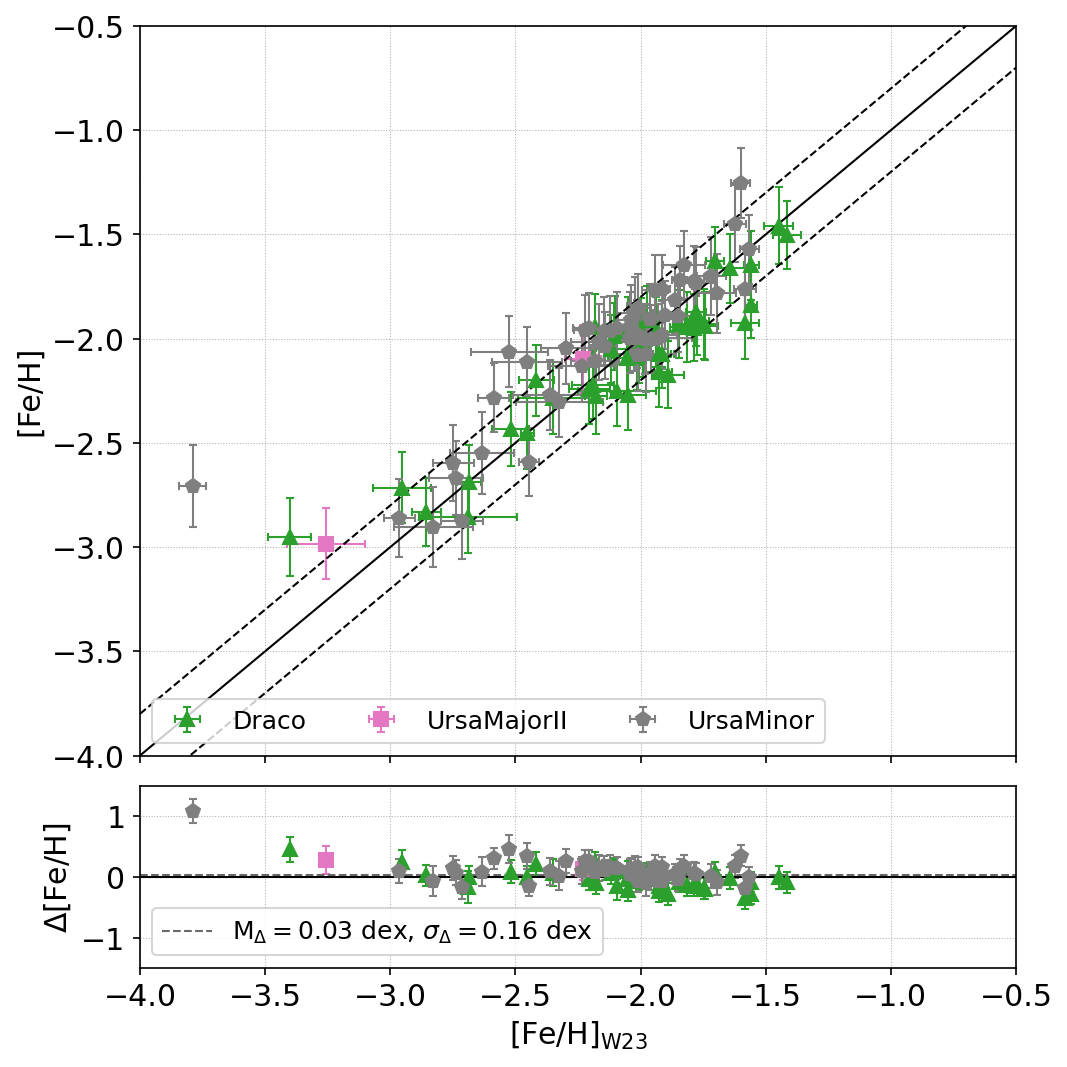}\\
    \includegraphics[width=.49\textwidth]{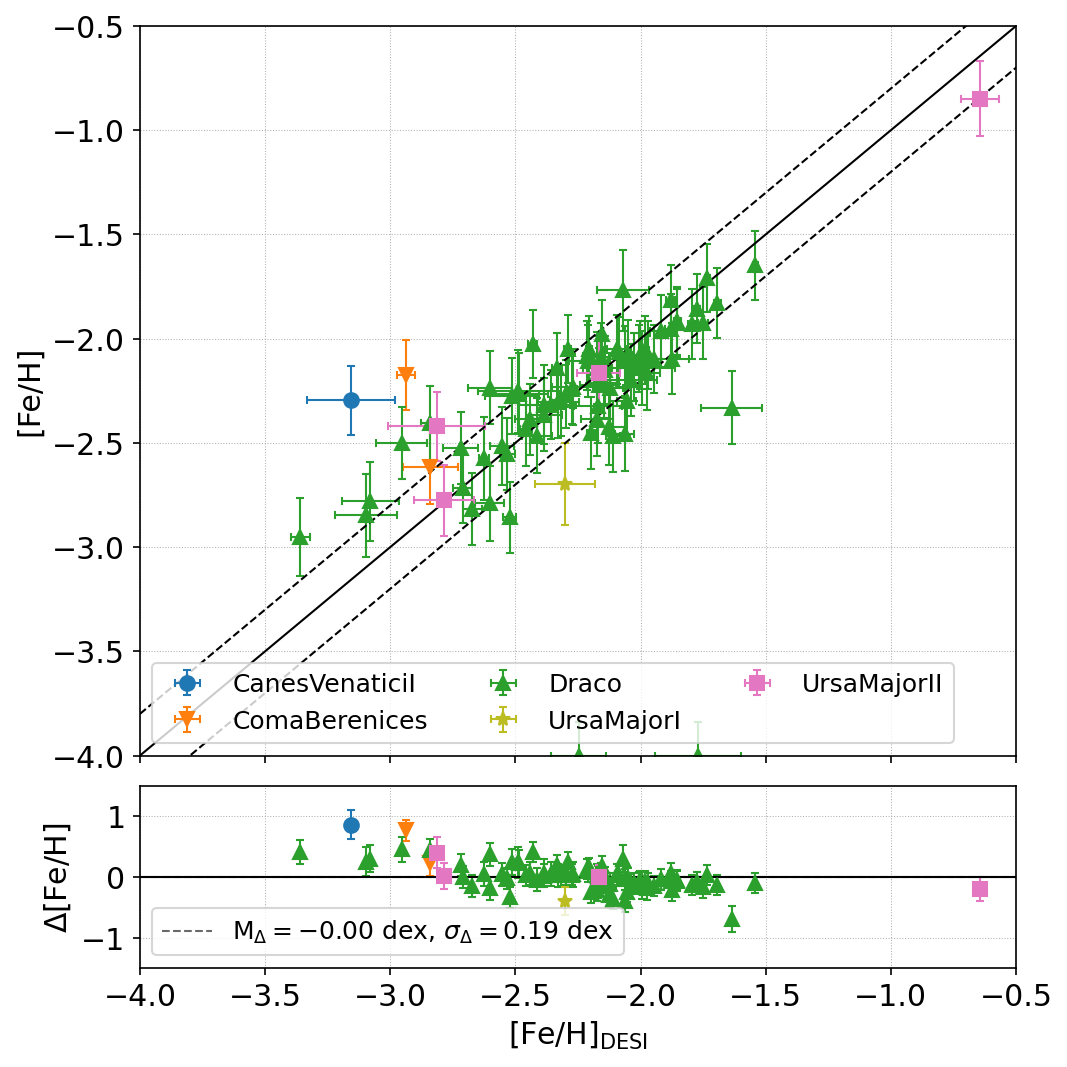}
    \includegraphics[width=.49\textwidth]{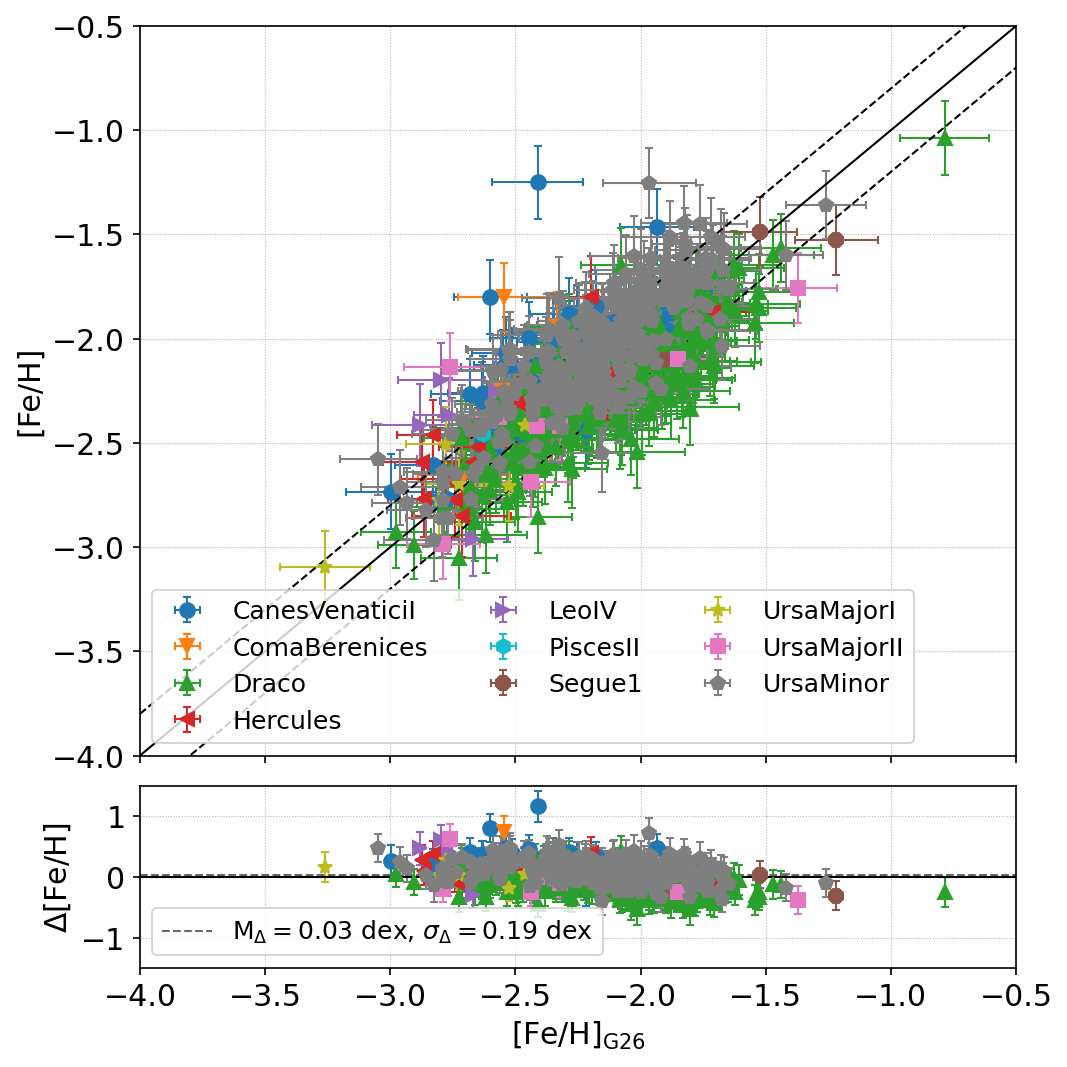}
    \caption{Metallicity validation using spectroscopic samples from the literature. The \textit{top left} panel shows the comparison with the high-resolution spectroscopic compilation reported in Sect.~\ref{apx:ext-comparison}. The \textit{top right} panel shows the comparison with the \citet{Walker2023} compilation. The bottom left panel shows the comparison with the \citet{Koposov2026} compilation. The bottom right panel shows the comparison with the \citet{Geha2026b} compilation. All panels have a dedicated inset for the residuals, with the median and scatter reported in the legend boxes.}
    \label{fig:validation-ext}
\end{figure*}

We compared the main photometric metallicity catalogue with a set of external catalogues.
In particular, we used high-resolution spectroscopic data ($R>25\,000$) of stars in dwarf galaxies from literature sources that were not included in SAGA and therefore not part of the calibration sample.

Data were collected from the following publications: \citet[][CBer, Seg1]{Sitnova2021}; \citet[][CVnI, Her]{Francois2016}; \citet[][Dra]{Cohen+Huang2009}; \citet[][Dra]{Tsujimoto2017}; \citet[][PscII]{Spite2018}; \citet[][UMaII]{Frebel2010}; and \citet[][UMi]{Cohen+Huang2010}. As in Sect.~\ref{apx:int-comparison}, we imposed quality cuts on the photometric metallicities and restricted the uncertainties to 0.2~dex.

As shown in Fig.~\ref{fig:validation-ext} (top left panel), the comparison with the literature data is excellent. Despite the heterogeneity of the sources, there are no significant shifts, with most of the values agreeing well within the errors. The scatter on the residuals is also comparable to that obtained with the calibration sample (see Fig.~\ref{fig:validation-int}). 

We can also see from the figure that the carbon-enhanced metal-poor stars (CEMP-no) identified in Seg1 have calculated ${\rm [Fe/H]_{Pr}}$ values that are more metal-rich than their spectroscopic values. This was expected since strong carbon bands can significantly bias the estimated photometric metallicity when using the CaHK filter, as shown in \citet{Martin2024}.

We further cross-matched our results with other spectroscopic compilations, namely those of \citet{Geha2026}, DESI-DR1 \citep{Koposov2026}, and \citet{Walker2023}. We note that the comparison with \citet{Walker2023} is not a completely independent test, as part of this data were included in the calibration sample. However, in this case, the cross-match extends to fainter sources with a match with the M18 photometry.

As for \citet{Walker2023}, also DESI-DR1 metallicities are on the \texttt{APOGEE} scale. Therefore, we subtracted 0.1~dex to bring them into agreement with the [Fe/H] scale used in this work. On the other hand, we did not apply any shift to the \citet{Geha2026} data.

Along with the quality cuts mentioned above, we only consider stars with a calculated membership probability $\mathcal{P}_{\rm i}\geq0.05$, a spectroscopic $\delta{\rm [Fe/H]}<0.2$~dex and a $S/N>10$ (when available). 
As shown in Fig.~\ref{fig:validation-ext}, there is excellent agreement between the photometric and spectroscopic values from \citet{Walker2023} and DESI-DR1, with median offsets and scatter comparable to previous cases. 

Similarly good results were obtained when comparing with the \citet{Geha2026} compilation, although small systematic offsets of $\sim 0.1$~dex were also evident on a system-by-system basis, with the notable exception of Leo~IV ($\sim 0.5$~dex). 
These results show that our photometric metallicities compare well in general to spectroscopic compilations covering a range of resolutions, from medium-high ($R>25,000$, as in Walker et al.) to medium-low ($R\lesssim5,000$, as in DESI-DR1 and Geha et al.).

\subsection{Metallicity properties of the individual systems}

\begin{figure*}
    \centering
    \includegraphics[width=.75\textwidth]{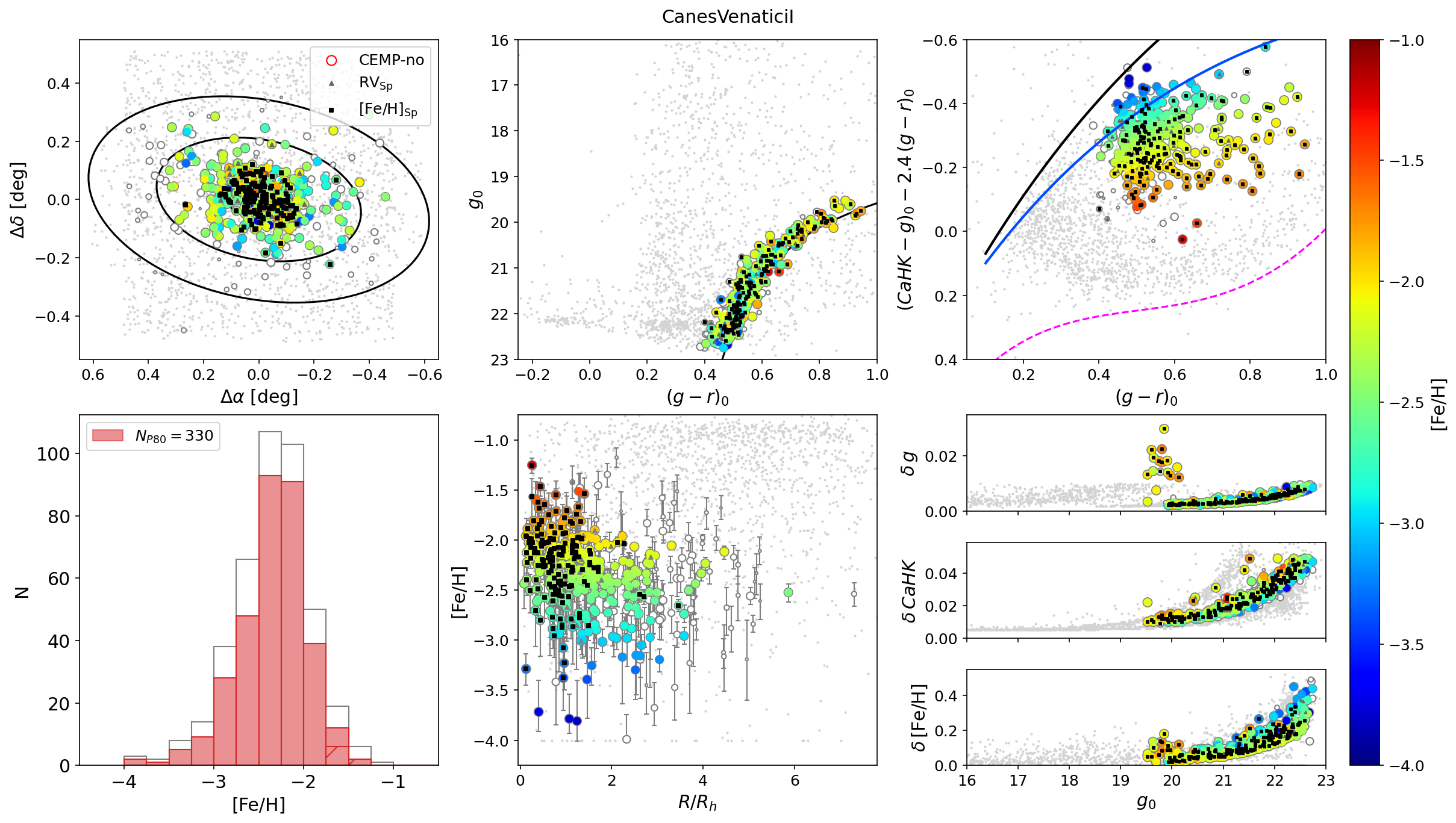}\\
    \includegraphics[width=.75\textwidth]{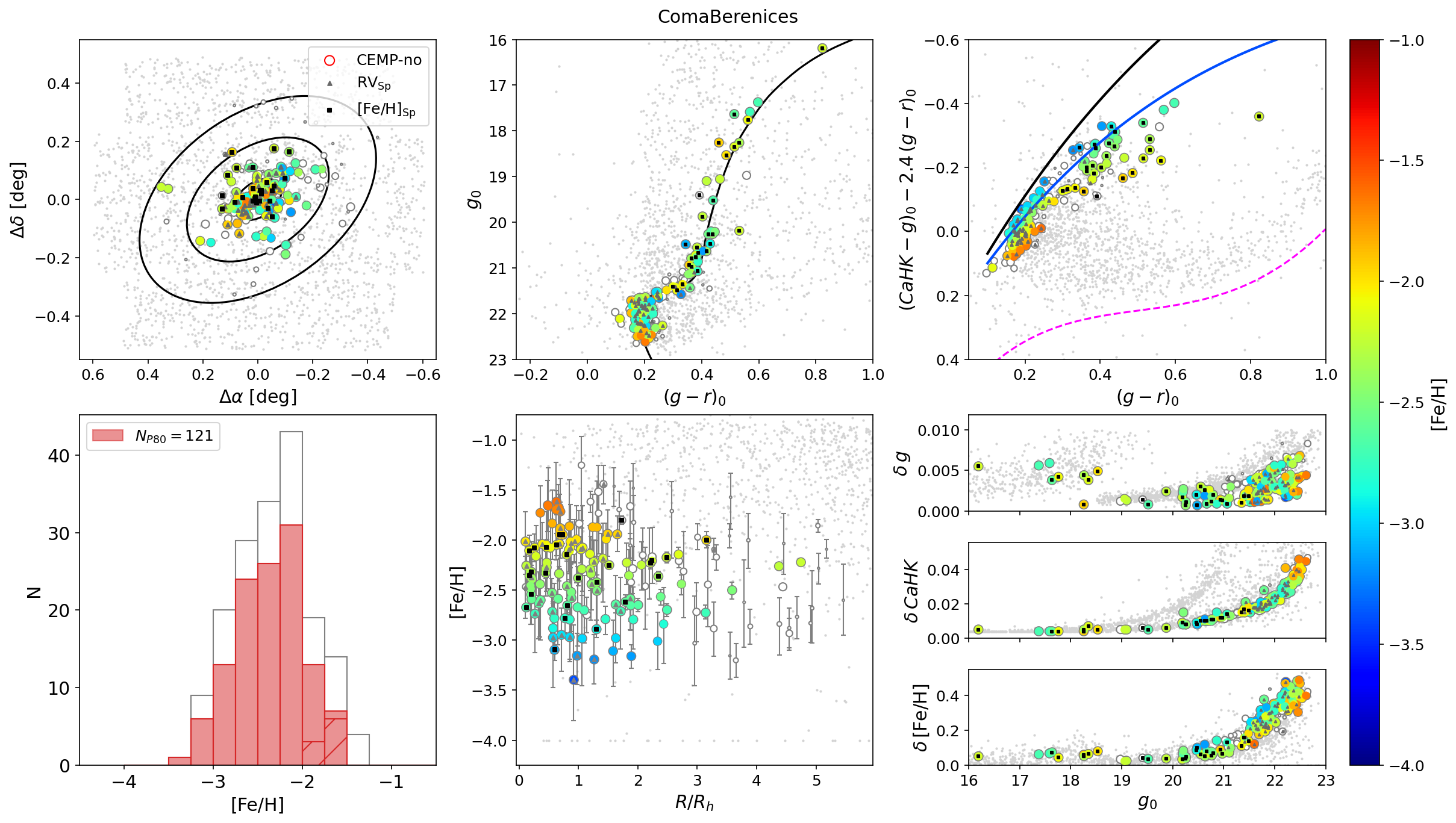}\\
    \includegraphics[width=.75\textwidth]{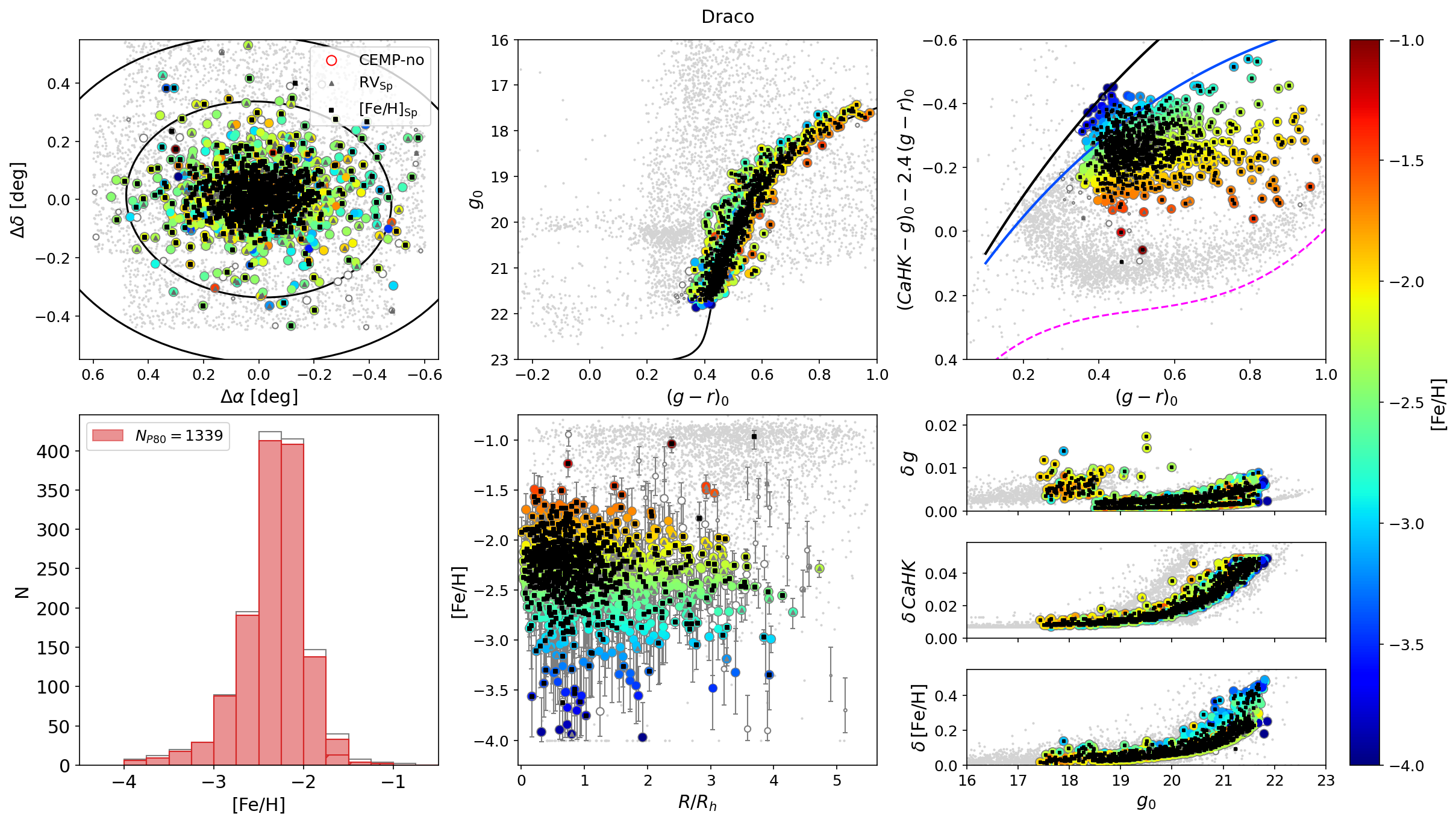}\\
    \caption{
    %Properties of the dwarf galaxy sample. 
    Panels from the top left show the probable members: spatial distribution with known spectroscopic targets and with ellipses at $1\times$, $3\times$ and $5\times R_h$ semi-major axes; CMD with 13~Gyr isochrone at nearest [Fe/H]; CaHK calibration plane with iso-metallicity lines; MDF of stars with $\mathcal{P}_{\rm f}\geq0.8$ and $\mathcal{P}_{\rm f}\geq0.05$; radial [Fe/H] distribution in $R_h$ units; magnitude and [Fe/H] uncertainty trends.
    }
    \label{fig:FEH-dist-UFD-1}
\end{figure*}

\begin{figure*}
    \centering
    \includegraphics[width=.75\textwidth]{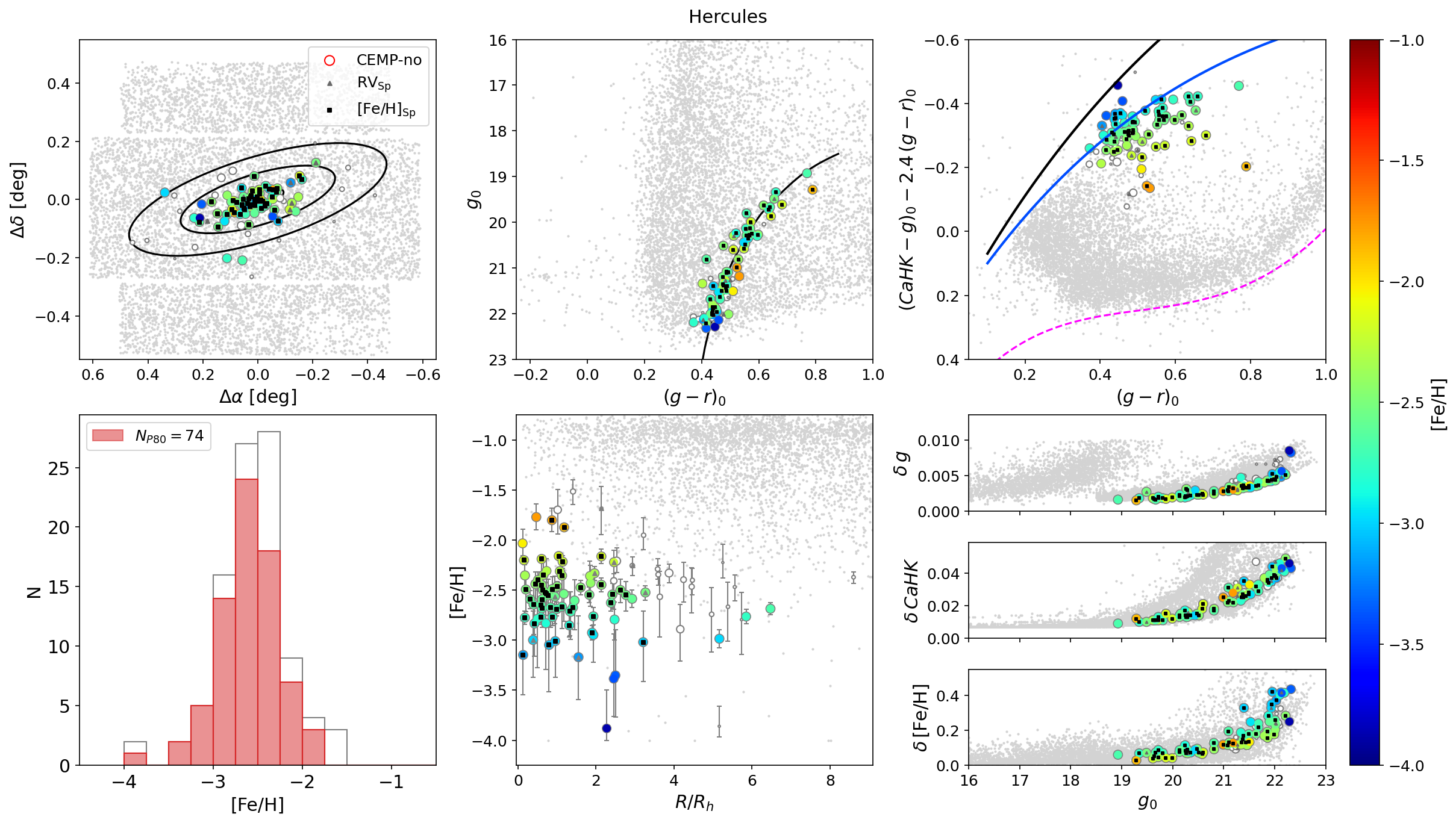}\\
    \includegraphics[width=.75\textwidth]{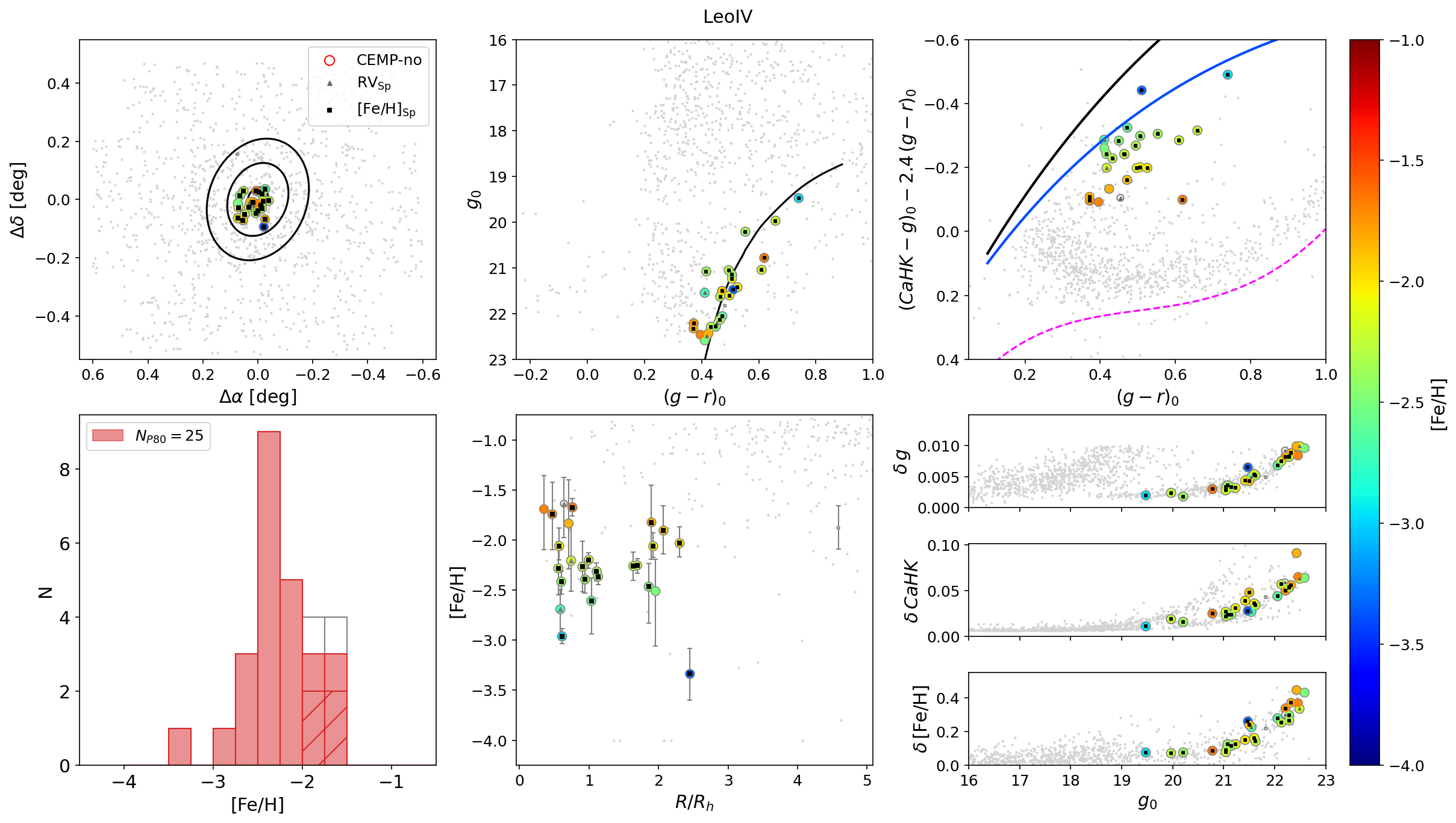}\\
    \includegraphics[width=.75\textwidth]{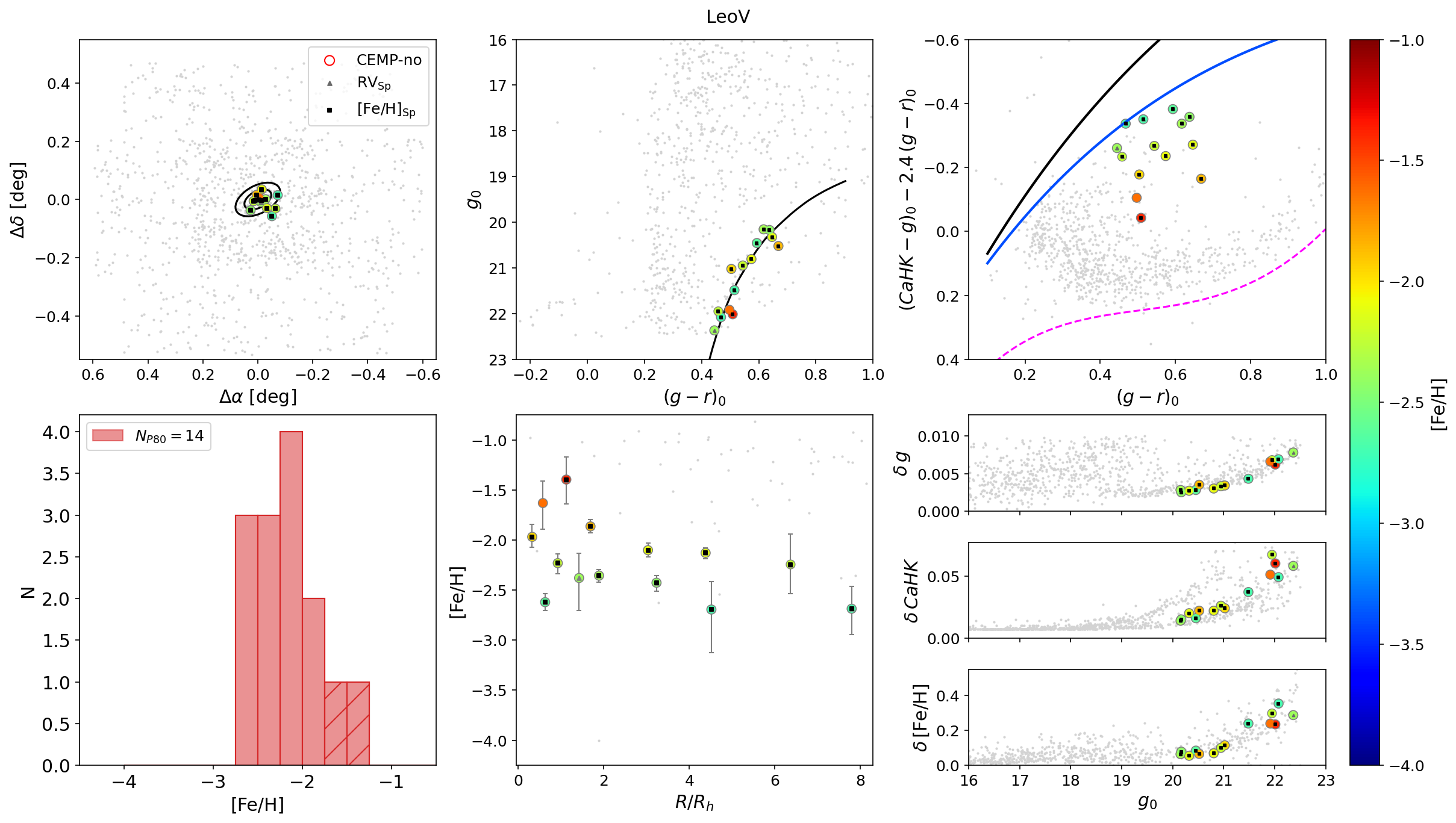}\\
    \caption{Continue from Fig.~\ref{fig:FEH-dist-UFD-1}.}
    \label{fig:FEH-dist-UFD-2}
\end{figure*}

\begin{figure*}
    \centering
    \includegraphics[width=.75\textwidth]{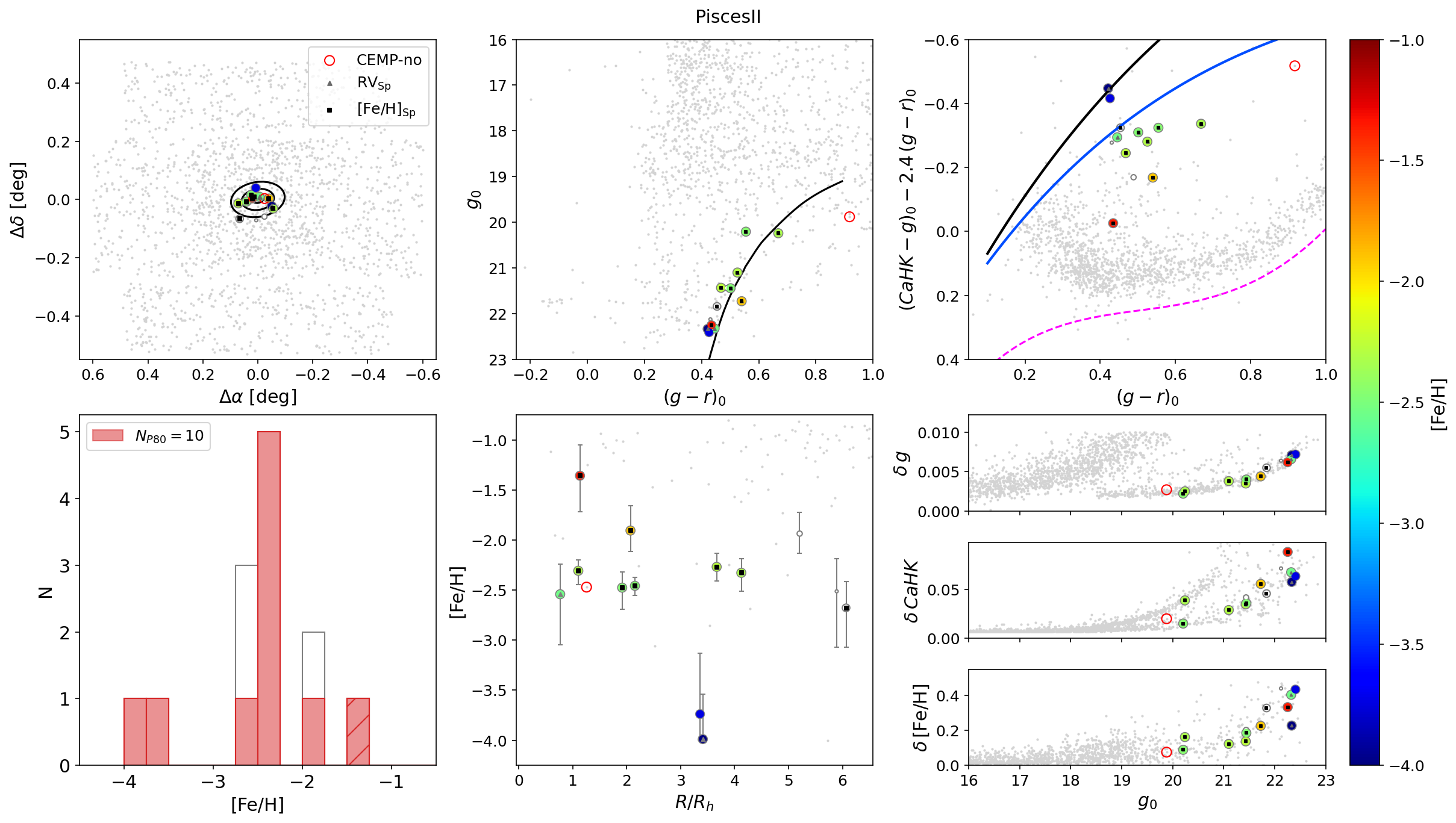}\\
    \includegraphics[width=.75\textwidth]{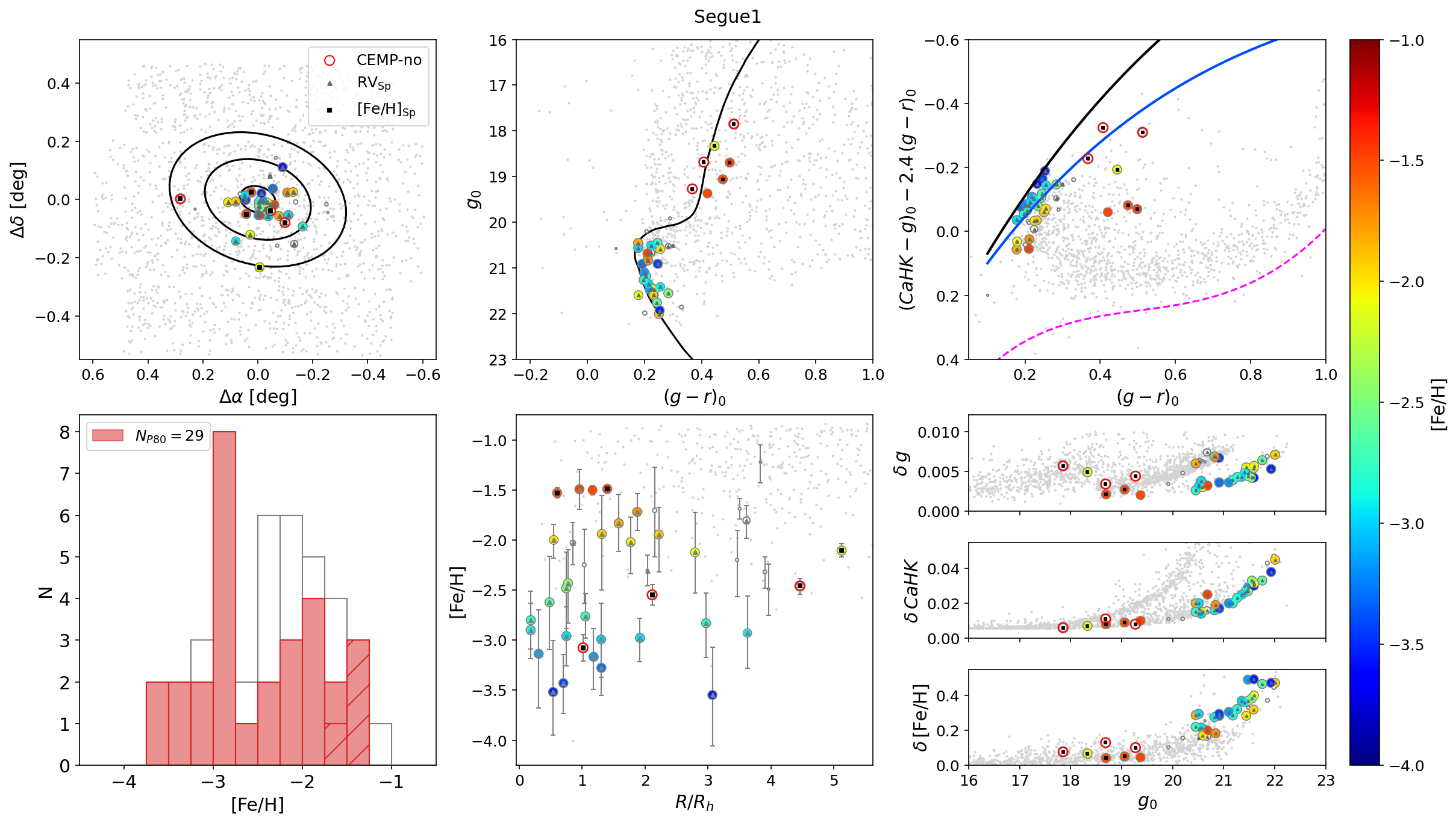}\\
    \includegraphics[width=.75\textwidth]{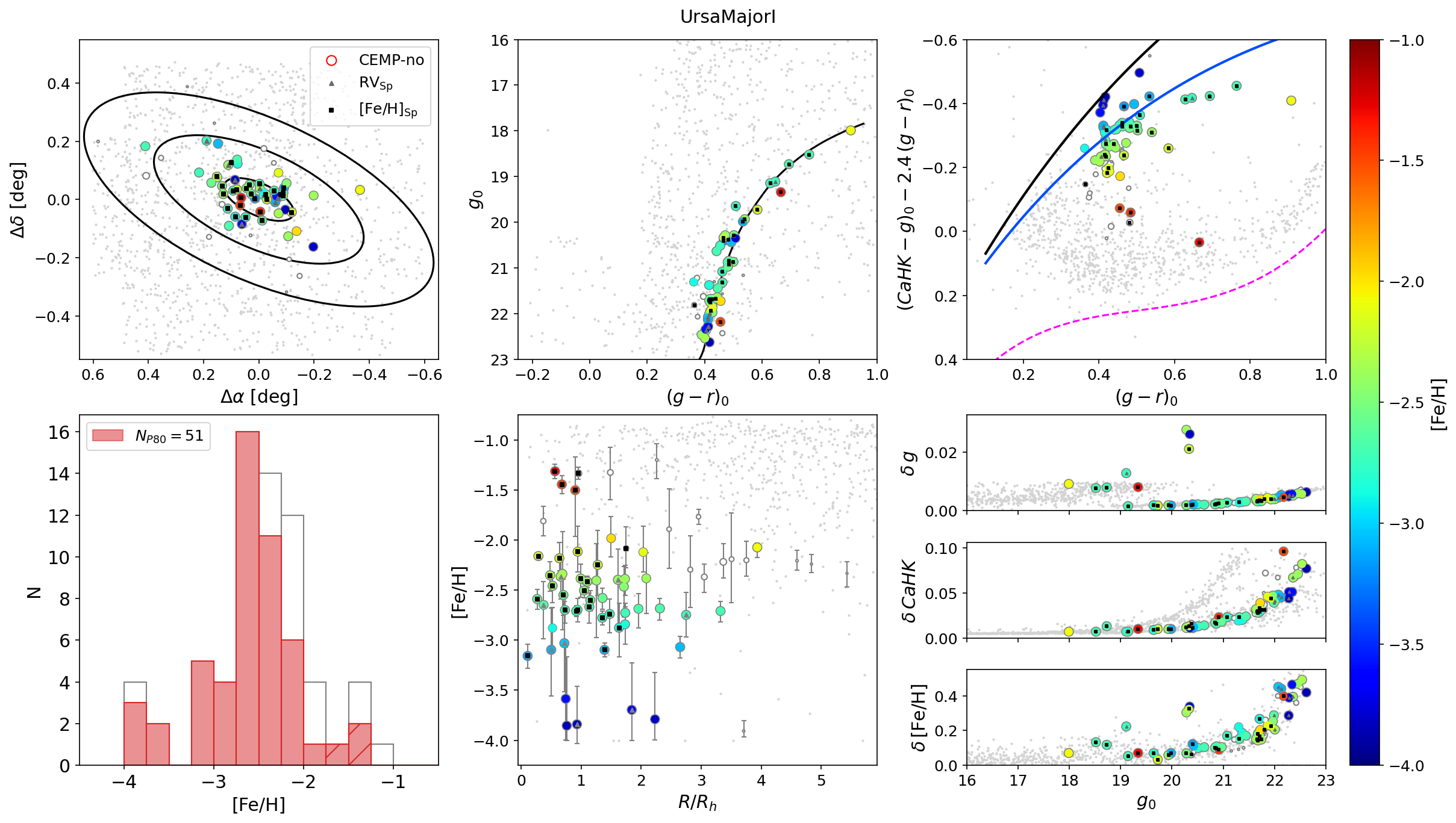}\\
    \caption{Continue from Fig.~\ref{fig:FEH-dist-UFD-1}.}
    \label{fig:FEH-dist-UFD-3}
\end{figure*}

\begin{figure*}
    \centering
    \includegraphics[width=.75\textwidth]{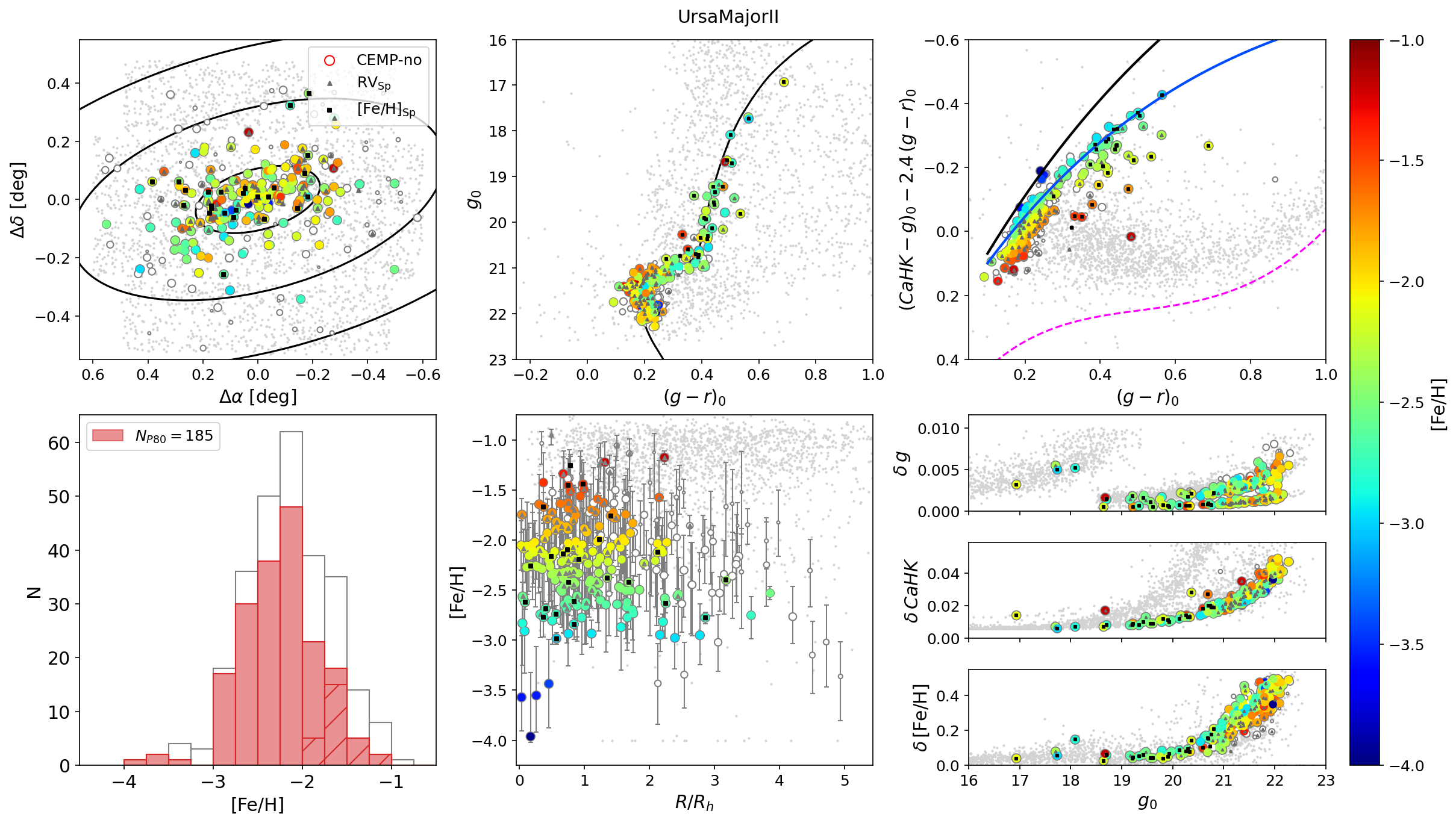}\\
    \includegraphics[width=.75\textwidth]{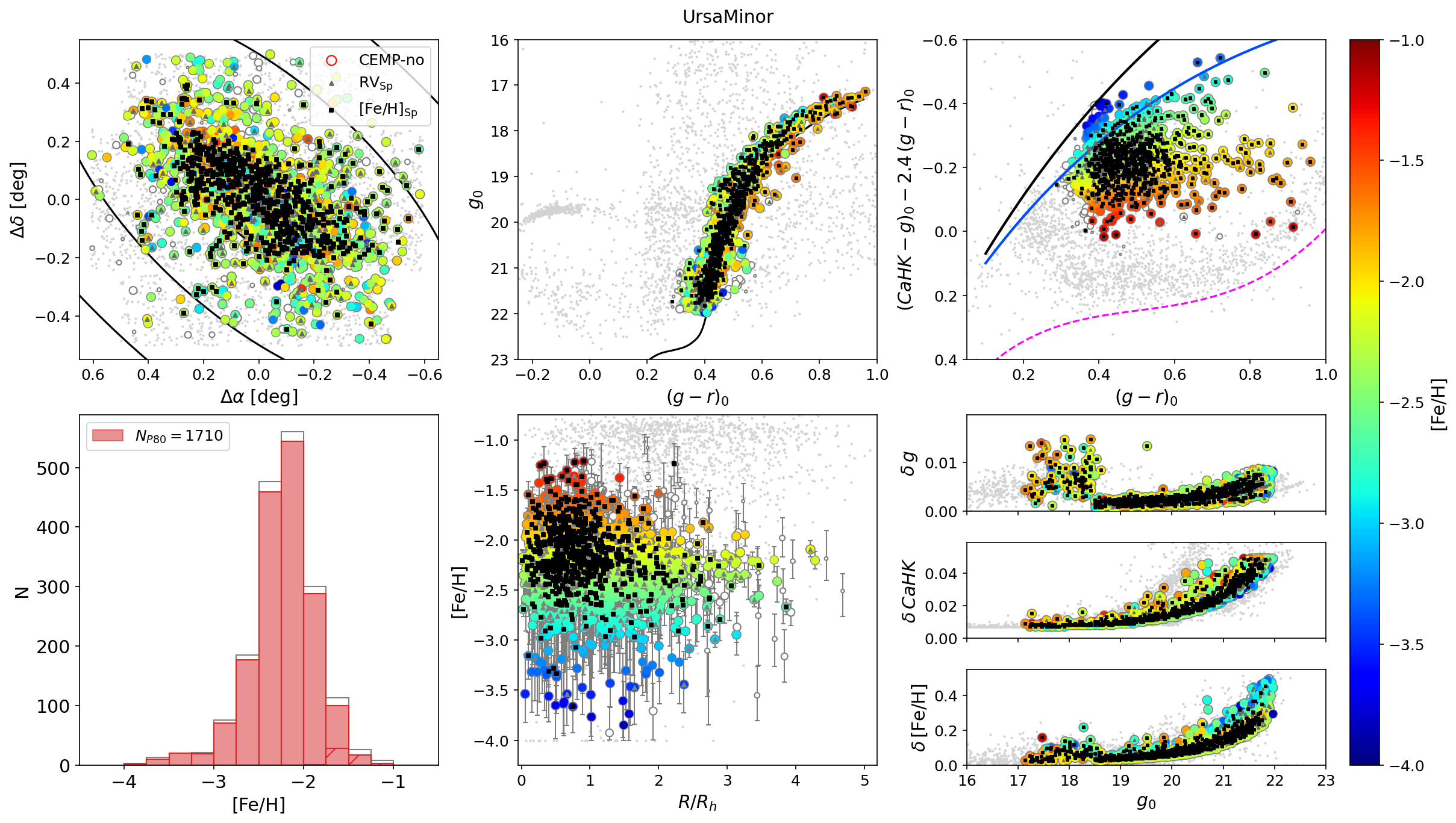}\\
    \includegraphics[width=.75\textwidth]{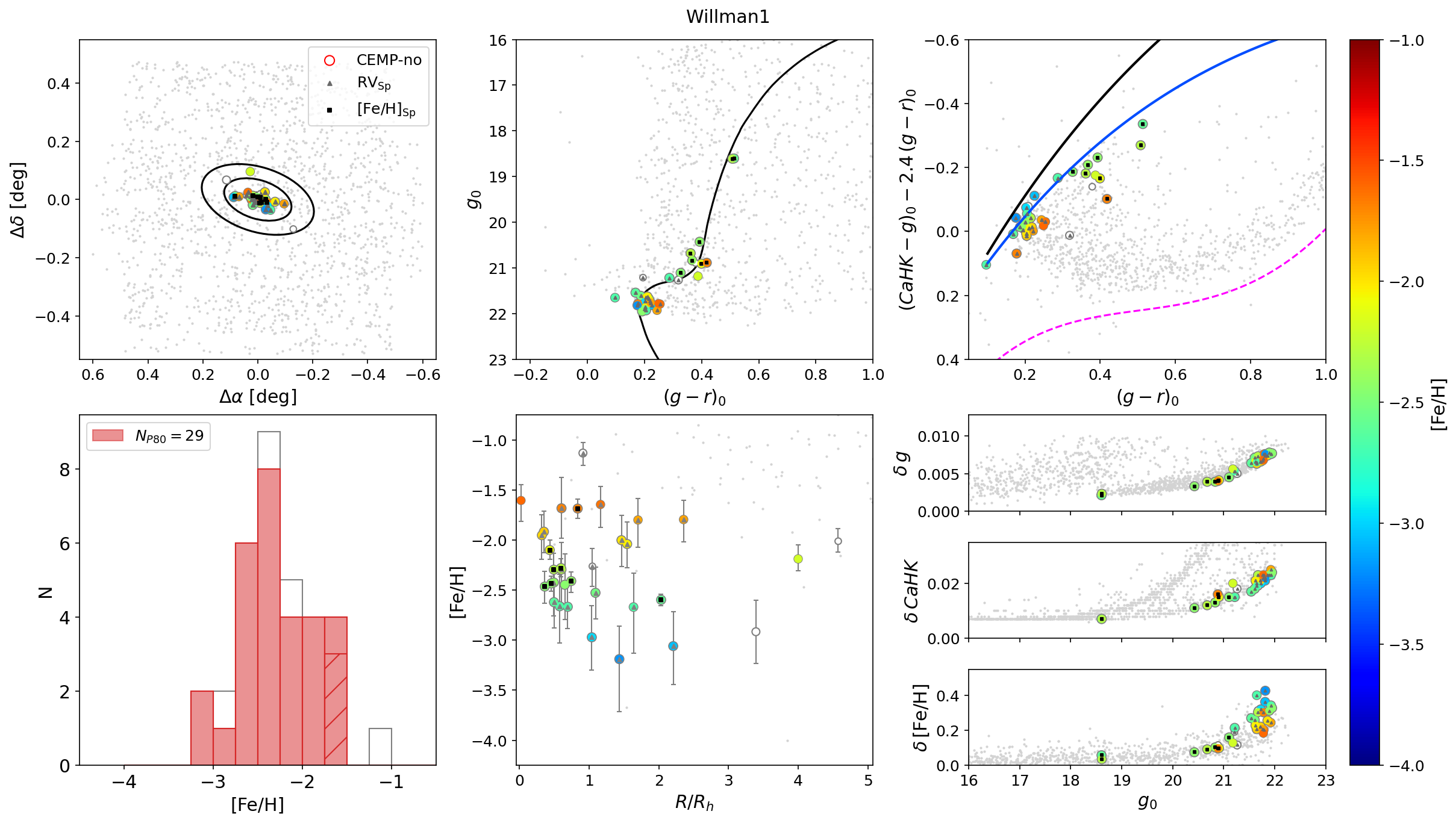}\\
    \caption{Continue from Fig.~\ref{fig:FEH-dist-UFD-1}.}
    \label{fig:FEH-dist-UFD-4}
\end{figure*}

The metallicity distribution properties of the dwarf galaxies in our sample are shown in Figs.~\ref{fig:FEH-dist-UFD-1}-\ref{fig:FEH-dist-UFD-4}. For each system, the figures show, in clockwise order starting from the top left panel, the probable members: spatial distribution, with ellipses having semi-major axes equal to $1\times$, $3\times$ and $5\times R_h$; colour-magnitude diagram with a 13~Gyr Dartmouth isochrone \citep{Dotter2008} with the closest metallicity to the systemic one; CaHK calibration plane with iso-metallicity lines as in Fig.~\ref{fig:grid}; metallicity distribution histograms of stars with $\mathcal{P}_{\rm f}\geq0.8$ and $\mathcal{P}_{\rm f}\geq0.05$; calculated metallicities and their associated uncertainties as a function of the semi-major axis radius in units of the half-light radius $R_h$; $g_0$, $CaHK_0$, and [Fe/H] uncertainties. Figures are colour-coded according to the calculated photometric metallicities. The black (gray) squares (triangles) indicate stars for which a spectroscopic metallicity (or radial velocity) measurement is available.

%%%%%%%%%%%%%%%%%%%%%%%%%%%%%%%%%%%%%%%%%%%%%%%%%%

% Don't change these lines
\bsp	% typesetting comment
\label{lastpage}
\end{document}